\documentclass[11pt]{article}

\usepackage[utf8]{inputenc}
\usepackage{amsmath}
\usepackage{amsthm}
\usepackage{amsfonts}
\usepackage{amssymb}
\usepackage{graphicx}
\usepackage{mathtools}
\usepackage{natbib}
\usepackage{enumitem}
\usepackage{url}
\usepackage{authblk}
\usepackage{bm}
\usepackage[usenames]{color}
\usepackage{hyperref}
\usepackage{geometry}
\usepackage{caption}
\usepackage{float}
\usepackage[caption = false]{subfig}
\usepackage{tikz}
\usepackage{multirow}
\usepackage[linesnumbered, ruled,vlined]{algorithm2e}
\usepackage{pdflscape}
\newcommand{\R}{\mathbb{R}}

\newcommand{\x}{\textbf{x}}

\def\mbf#1{\mathbf{#1}} % bold but not italic
\newcommand{\simiid}{\stackrel{iid}{\sim}} %[] IID 
\def\where{\text{ where }} % where
\newcommand{\indep}{\perp \!\!\! \perp } % independent symbols
\def\mrm#1{\mathrm{#1}} % remove math
\newcommand{\reals}{\mathbb{R}} % Real number symbol
\def\t#1{\tilde{#1}} % tilde
\def\normal#1#2{\mathcal{N}(#1,#2)} % normal
\def\mc#1{\mathcal{#1}} % mathical
\DeclareMathOperator*{\argmin}{arg\,min} % arg min
\def\E#1{\mathrm{E}(#1)} % Expectation symbol
\def\var#1{\mathrm{Var}(#1)} % Variance symbol
\newcommand{\norm}[1]{\left\lVert#1\right\rVert} % A norm with 1 argument
\DeclareMathOperator{\Var}{Var} % Variance symbol

\newtheorem{lem}{Lemma}

\newtheorem{prop}{Proposition}
\theoremstyle{definition}
\newtheorem{remark}{Remark}
\hypersetup{
  linkcolor  = blue,
  citecolor  = blue,
  urlcolor   = blue,
  colorlinks = true,
} % color setup

\newenvironment{proof-of-proposition}[1][{}]{\noindent{\bf
    Proof of Proposition {#1}}
  \hspace*{.5em}}{\qed\bigskip\\}
  \newenvironment{proof-of-corollary}[1][{}]{\noindent{\bf
    Proof of Corollary {#1}}
  \hspace*{.5em}}{\qed\bigskip\\}
  \newenvironment{proof-of-lemma}[1][{}]{\noindent{\bf
    Proof of Lemma {#1}}
  \hspace*{.5em}}{\qed\bigskip\\}

\allowdisplaybreaks

\title{Minimizing post-shock forecasting error through aggregation of outside information}
\author{Jilei Lin\thanks{jileil2@ilinois.edu} }
\author{Daniel J. Eck\thanks{dje13@illinois.edu}}
\affil{Department of Statistics, University of Illinois at Urbana-Champaign}

\makeatletter
\long\def\@makecaption#1#2{
  \vskip 0.8ex
  \setbox\@tempboxa\hbox{\small {\bf #1:} #2}
  \parindent 1.5em  %% How can we use the global value of this???
  \dimen0=\hsize
  \advance\dimen0 by -3em
  \ifdim \wd\@tempboxa >\dimen0
  \hbox to \hsize{
    \parindent 0em
    \hfil 
    \parbox{\dimen0}{\def\baselinestretch{0.96}\small
      {\bf #1.} #2
    } 
    \hfil}
  \else \hbox to \hsize{\hfil \box\@tempboxa \hfil}
  \fi
}
\makeatother

\begin{document}

\maketitle
\begin{abstract}
    We develop a forecasting methodology for providing credible forecasts for time series that have recently undergone a shock. We achieve this by borrowing knowledge from other time series that have undergone similar shocks for which post-shock outcomes are observed. Three shock effect estimators are motivated with the aim of minimizing average forecast risk. We propose risk-reduction propositions that provide conditions that establish when our methodology works. Bootstrap and leave-one-out cross validation procedures are provided to prospectively assess the performance of our methodology. Several simulated data examples, and a real data example of forecasting Conoco Phillips stock price are provided for verification and illustration.
\end{abstract}

\section{Introduction}

We provide forecasting adjustment techniques with the goal of lowering overall forecast error when the time series under study has undergone a structural shock. We focus on the specific setting in which a structural shock has occurred and one desires a prediction for the post-shock response at the next time point. Standard forecasting methods may not yield accurate predictions in the presence of such structural shocks \citep{baumeister2014real}. This is a general problem that has many real life applications. For example, one may be interested in forecasting the stock price of a company tomorrow after hearing terrible or great news about the company  after hours trading. Companies may be interested in forecasting the demand of their products to adjust production after they were involved in a brand crisis, but they only have recent sales data for which the company is operating well. All is not lost in this setting, one may be able to supplement the present forecast with past data borrowed from other time series which contain similar structural shocks. The core idea of our methodology is to sensibly aggregate similar past realized shock effects which arose from other time series, and then incorporate the aggregated shock effect estimator into the present forecast. Our method of combining shock effects embraces ideas from conditional forecasting \citep{baumeister2014real, kilian2017structural}, time series pooling using cross-sectional panel data \citep{ramaswamy1993empirical, pesaran1999pooled, hoogstrate2000pooling, baltagi2008forecasting, koop2012forecasting, liu2020forecasting}, forecasting with judgement and models \citep{svensson2005monetary, monti2008forecast}, synthetic control methodology \citep{abadie2010synthetic, agarwal2020two}, and expectation shocks \citep{croushore2006data, baumeister2014general, clements2019measuring}. 

We study the post-shock forecasting problem in the context of additive shock effects in linear autoregressive models. In this post-shock forecasting setting, the researcher has a time series of interest which is known to have recently undergone a structural shock, and the post-shock response is not observed. In this setting, the additive shock effect is a random effect that is parameterized in the autoregressive model. The shock effect is then estimated using ordinary least squares (OLS). The researcher must move beyond the modeling paradigm that they were previously working under to accommodate this new shock effect \citep{monti2008forecast, svensson2005monetary}. One method for estimating the shock effect is to produce a conditional point forecast where a sequence of non-zero future structural shocks are conditioned upon and estimated \citep{baumeister2014real}. Such conditional point forecasts are appropriate when the shock sequence considered is within the range of historical experience \citep{kilian2017structural}. On the other hand, our methodology allows for the inclusion of outside data sources and covariates into this conditional forecasting context provided that the shock effects from outside data sources are all thought to arise from a  data generating process similar to that of the shock under study. For further differences of assumptions on shocks, our methodology allows for unprecedented shocks and no observation of past shocks. 

In our methodological framework, the researcher creates a synthetic panel of time series which have undergone similar structural shocks in the past. Construction of the donor pool that forms this synthetic panel is similar to that in synthetic control methodology (SCM) \citep{abadie2010synthetic}. As in SCM, care is needed when forming the donor pool of time series. However, there are key differences between our framework and SCM. We assume that the time series forming the donor pool are independent from the time series under study before the timing of the shock. Moreover, the shocks from candidate time series in the donor pool, together with the shock in the time series of interest, are assumed to be  from a common family of distributions with existing first and second moments. 

We estimate the shock effects that are present in the time series forming the donor pool for which post-shock responses are observed. We then aggregate these estimated shock effects and use this aggregated estimate as an estimator for the shock effect in the time series of interest. This estimator is then added to a forecast for the yet to be realized post-shock response corresponding to the time series of interest. Shock effects in our post-shock forecasting framework is similar to ``expectation shocks'' which are studied in \cite{clements2019measuring}. The context in \cite{clements2019measuring} allows for consistent estimation of expectation shocks under a vector autoregressive model, possibly involving an instrumental variable approach as in \cite{croushore2006data}. In our context, the yet to be observed shock effect of interest is a random effect, and we can only partially estimate features of the random effect distribution using the time series forming the donor pool.

%Therefore the traditional forecast combination framework may not be of any help.  
%The Bayesian hierarchical approach of \citet{lee2020estimation} may be too sensitive to prior and the hierarchical parametric model setup. Overall, it is very unlikely that the above mentioned methods will work ideally since they are trained on the time series data that do not experience such a shock. 

In this article, we will assume a simple autoregressive data generating process similar to that in \citet{blundell1998initial} with a general random-effect structure. Therefore, our methodology is similar to the ``$K$ latent pooling'' framework of \cite{ramaswamy1993empirical}. However, our model formulation is more general than \cite{ramaswamy1993empirical}. In our model, the donor pool can consist of dependent time series but time series within the donor pool should be independent of the time series of interest. However, mutual independence among time series in the donor pool can aid prospective evaluation of the reliability of our method. We consider three aggregation techniques: simple averaging, inverse-variance weighted averaging, and similarity weighting. The latter technique is similar to the weighting in synthetic control methodology \citep{abadie2010synthetic}. Our autoregressive model will consider present day covariates to better motivate similarity weighting. The considered adjustment strategies all target the mean of the shock effect distribution. Such an estimation strategy can reduce mean squared error (MSE) when variation in the shock effect distribution is small relative to the mean. We provide risk-reduction propositions that detail the conditions when the adjusted forecasts will work better than the original forecast. The involved parameters in the risk-reduction propositions can be estimated by a residual bootstrap procedure that we develop. We also motivate a simple leave-one-out cross validation procedure which can prospectively assess the performance of our shock effect adjustment estimators. This prospective assessment does not require the observation of the post-shock response. Our Monte Carlo simulation results show that the risk-reduction propositions are nearly perfectly correct when the model for the shock effects is identified well with appropriate covariates under a fixed design. We demonstrate the utility of our methodology in a real data analysis in which we forecast the stock price of Conoco Phillips shares that experienced a large structural shock on March 9th, 2020. We will show that our proposed adjustment estimators yield much better results than no adjustment in this setting. We also use this example to demonstrate settings in which the shock effect may be decomposed into separate estimable parts. We now motivate our framework for post-shock forecasting.

% Note that this methodology is not motivated with the goal of unbiased, asymptotically unbiased, or consistent estimation for the shock effect of the time series under study.

%The question is then how to estimate the shock effects and improve the 
%prediction? This paper proposes three estimators, the adjustment estimator, weighted adjustment estimator, and inverse-variance weighted estimators. Section (to be updated) discusses properties of those estimators. Section (to be updated) compares the  interplay between prediction risk and different shock effects estimators. Section (to be updated) conducts simulation to justify our claims and certain properties that cannot be found analytically.

\begin{figure}[H]
  \begin{center}
    \subfloat{\includegraphics[height = 7cm]{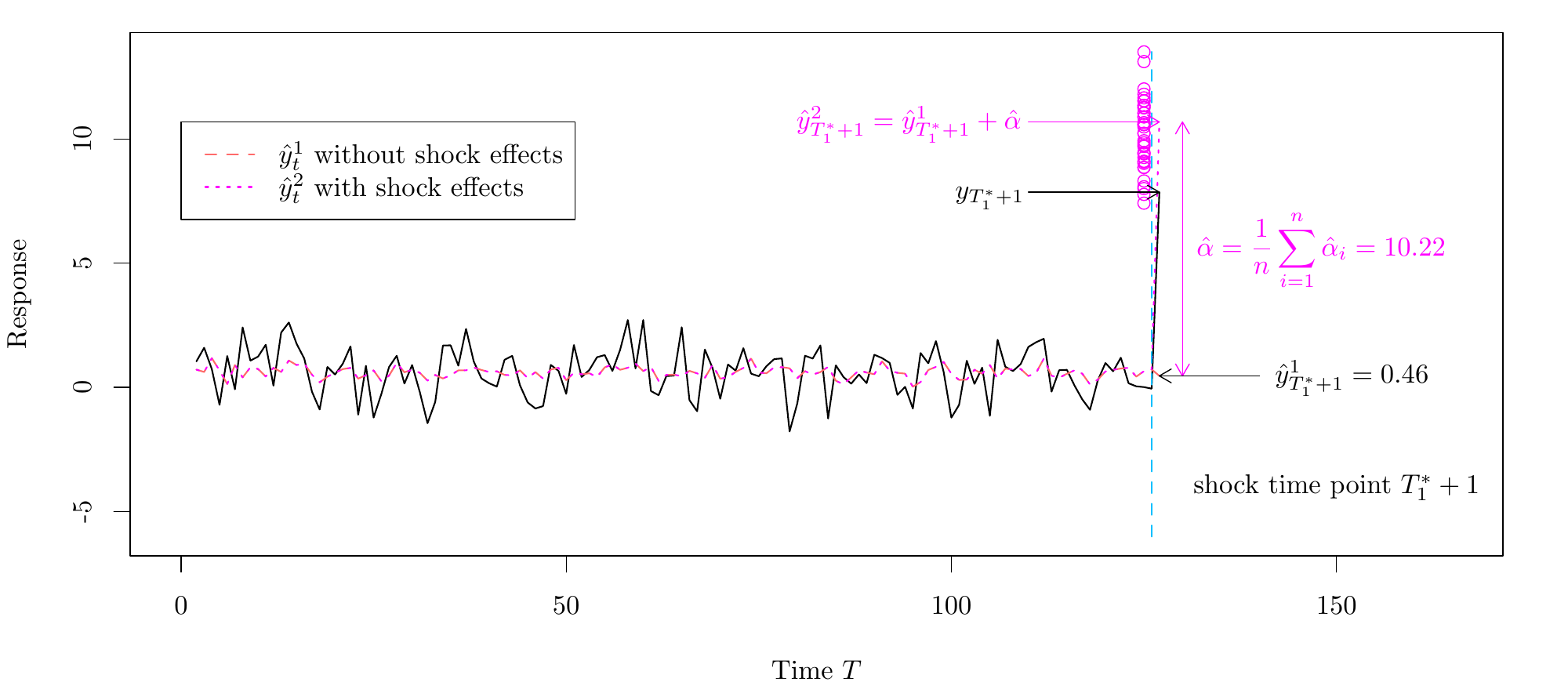}}\\
     \subfloat{\includegraphics[height = 7cm]{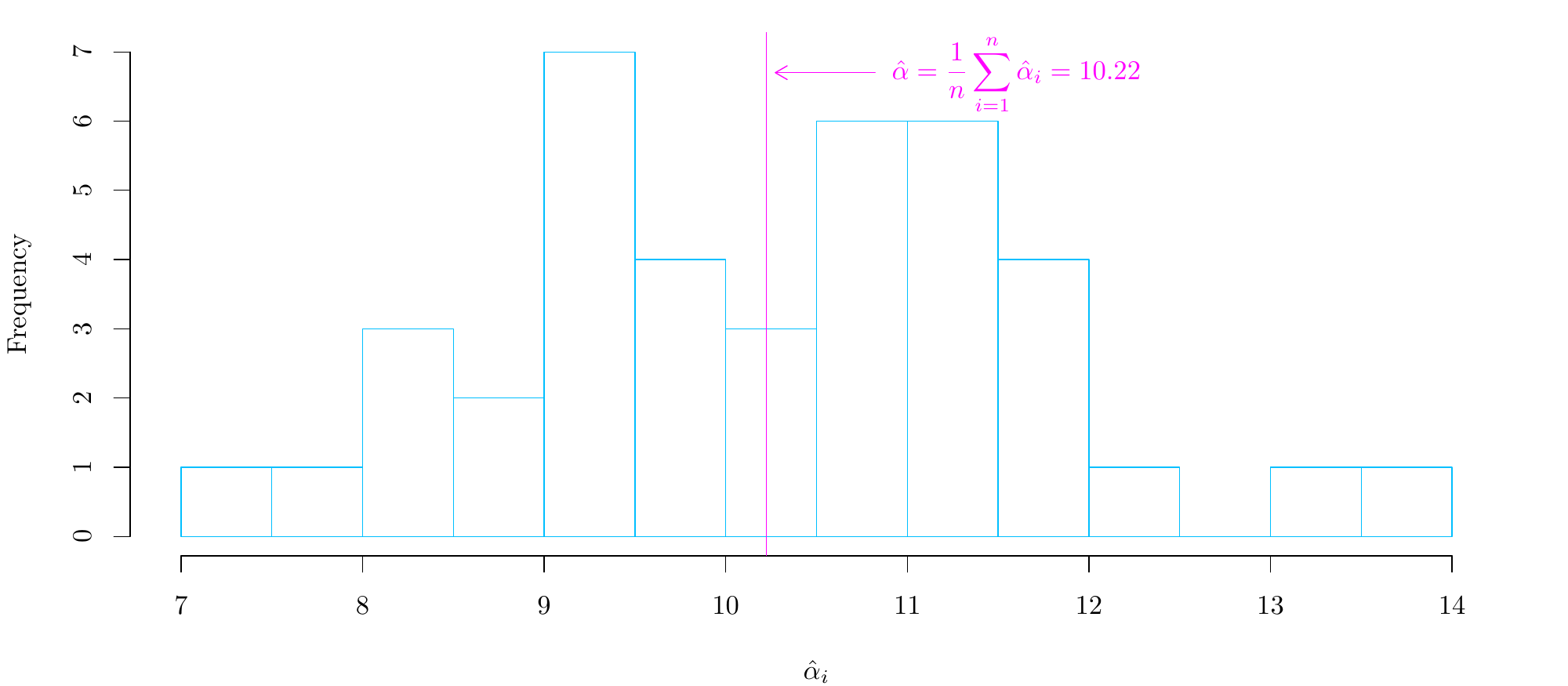}}
  \end{center}
     \caption{The time series experience a shock at $T_1^*+1=126$ with true shock effect $\alpha = 9.21$. (a) presents the comparison of the prediction without adjustment and one that uses simple averaging of estimated shock effects given a donor pool of $n=40$ time series. (b) shows the histogram for the least square estimates $\hat{\alpha}_i$s used in estimating shock $\alpha_1$ for $i = 2, \ldots, 41$.  The magenta dots represent $\hat{\alpha}_i$ from the time series forming the donor pool. The prediction of $\hat{y}^{2}_{T_1^*+1}$ and $\hat{y}^{1}_{T_1^*+1}$ differs only by an adjustment $\hat{\alpha}=10.22$. It is clear that $\hat{y}^{2}_{T_1^*+1}$ performs better than $\hat{y}^{1}_{T_1^*+1}$.}\label{figure1}
\end{figure}

\section{Setting}
\label{setting}

We will suppose that a researcher has time series data ($y_{i,t}$,$\x_{i,t}$), for $t = 1, \ldots,  T_i$ and $i = 1, \ldots, n+1$, where $y_{i,t}$ is a scalar response and $\x_{i,t}$ is a vector of covariates that are revealed to the analyst prior to the observation of $y_{1,t}$.  Suppose that the analyst is interested in forecasting $y_{1,t}$, the first time series in the collection. We will suppose that specific interest is in forecasting the response after the occurrence of a structural shock. To gauge the performance of forecasts, we consider forecast risk in the form of MSE,
$$
  R_T = \frac{1}{T}\sum_{t=1}^T\E{\hat y_{1,t} - y_{1,t}}^2,
$$
and root mean squared error (RMSE), given by $\sqrt{R_T}$, in our analyses. In this article, we focus on post-shock prediction where forecasts methods only differ at the next future time point. Thus the MSE reduces to the magnitude $\E{\hat y_{1,t} - y_{1,t}}^2$.

%The forecast consists of three steps: (1) pick a model, (2) selection of covariates, (3) choices of donor pool.
Our post-shock forecasting methodology will consist of selecting covariates $\x_{i,t}$, constructing a suitable donor pool of candidate time series that have undergone similar structural shocks to the time series under study, and specifying a model for the time series ($y_{i,t}$,$\x_{i,t}$), for $t = 1, \ldots,  T_i$ and $i = 1, \ldots, n+1$. In this article, we consider a dynamic panel data model with autoregressive structure similar to that in \citet{blundell1998initial}. Our dynamic panel model includes an additional shock effect whose presence or absence is given by the binary variable $D_{i,t}$, and we will assume that the donor pool time series are independent of the time series under study. The details of this model are in the next section.

Figure \ref{figure1} provides a simple intuition of the practical usefulness of our proposed methodology. This figure depicts a time series that experienced a shock at time point $T_1^*+1 = 126$. It is supposed that the researcher does not have any information beyond $T_1^*+1$, but does have observations from a donor pool of forty time series that have previously undergone a similar shock for which post-shock responses are recorded. Similarity in this context means that the shock effects are random variables that from a common distribution. In this example, the mean of the estimated shock effects is taken as a shock effect estimator for the time series under study. Forecasts are then made by adding this shock effect estimator to the estimated response values obtained from the estimation procedure that ignores the shock. It is apparent from Figure \ref{figure1} that adjusting forecasts in this manner 1) leads to a reduction in forecasting risk; 2) does not fully recover the true shock effect. We evaluate the performance of this post-shock forecasting methodology throughout this article; we outline situations for when it is expected to work and when it is not.

%Section \ref{properties} will provide a more detailed treatment about when our proposed estimators in Section \ref{constructionofestimators} will improve the prediction under various model setups in Section \ref{modelsetup}. More examples will be provided using Monte Carlo simulations in Section \ref{simulation}.

\subsection{Model Setup}

\label{modelsetup}

In this section, we will describe the assumed dynamic panel models for which 
post-shock aggregated estimators are provided. The basic structures of these models 
are the same for all time-series in the analysis, the differences between them lie in the setup of the shock effect distribution.

Let $I(\cdot)$ be an indicator function, $T_i$ be the time length of the time series $i$ for $i = 1, \ldots, n+1$, and $T_i^*$ be the time point just before the one when the shock is known to occur, with $T_i^* < T_i$.  For $t= 1, \ldots, T_i$ and $i = 1, \ldots, n+1$, the model $\mc{M}_1$ is defined as
\begin{align}
\mc{M}_1 \colon y_{i,t} =\eta_i +\alpha_i D_{i,t} + \phi_i y_{i, t-1} + \theta_i'\mbf{x}_{i,t} + \varepsilon_{i,t}\label{equation1}
\end{align}
 where $D_{i,t} = I(t = T_i^* + 1)$ 
and $\x_{i,t} \in \R^{p}$ with $p \geq 1$.  We assume that the 
$\mbf{x}_{i,t}$'s are fixed. Let $|x|$ denote the absolute value of $x$ for $x\in \reals$. For $i = 1, \ldots, n+1$ and $t=1, \ldots, T_i$, the random effects structure for $\mc{M}_1$ is:
\begin{align*}
  \eta_i &\simiid \mc{F}_{\eta} \text{ with }  \; \mrm{E}_{\mc{F}_{\eta}}(\eta_i) = 0, \mrm{Var}_{\mc{F}_{\eta}}(\eta_i)  = \sigma^2_{\eta}\\
  \phi_i &\simiid \mc{F}_{\phi} \text{ where } |\mc{F}_{\phi}| < 1, \\
   \theta_i &\simiid \mc{F}_{\theta} \text{ with }  \; \mrm{E}_{\mc{F}_{\theta}}(\theta_i) = \mu_{\theta}, \mrm{Var}_{\mc{F}_{\theta}}(\theta_i)  = \Sigma^2_{\theta} \\
\alpha_i &\simiid \mc{F}_{\alpha} \text{ with }  \; \mrm{E}_{\mc{F}_{\alpha}}(\alpha_i) = \mu_{\alpha}, \mrm{Var}_{\mc{F}_{\alpha}}(\alpha_i)  = \sigma^2_{\alpha}  \\
\varepsilon_{i,t} & \simiid  \mc{F}_{\varepsilon} \text{ with }  \; \mrm{E}_{\mc{F}_{\varepsilon}}(\varepsilon_{i,t}) = 0, \mrm{Var}_{\mc{F}_{\varepsilon}}(\varepsilon_{i,t})  = \sigma^2 \where \sigma > 0 ,  \\
\eta_i &\indep  \alpha_i \indep \phi_i \indep \theta_i \indep \varepsilon_{i,t}.
\end{align*}
Notice that $\mc{M}_1$ assumes that $\alpha_i$ are iid with $\E{\alpha_i}=\mu_{\alpha}$ 
for $i = 1, \ldots, n+1$. We also consider a model where the shock effects are linear functions of covariates with an additional additive mean-zero error. For $i = 1, \ldots, n+1$, the random effects structure for this model (model $\mc{M}_2$) is:
\begin{align}
\mc{M}_2 \colon \begin{array}{l}
  y_{i,t} =\eta_i +\alpha_i D_{i,t} + \phi_i y_{i, t-1} + \theta_i'\mbf{x}_{i,t} + \varepsilon_{i,t}\\[.2cm]
  \; \alpha_i = \mu_{\alpha}+\delta_{i}'\mbf{x}_{i, T_i^*+1}+ \t{\varepsilon}_{i},
\end{array}\label{model2}
\end{align}
 where the added random effects are
\begin{align*}
\t{\varepsilon}_{i} &\simiid  \mc{F}_{\t{\varepsilon}} \text{ with }\mrm{E}_{\mc{F}_{\t{\varepsilon}}}(\t{\varepsilon}_{i})=0, \mrm{Var}_{\mc{F}_{\t{\varepsilon}}}(\t{\varepsilon}_{i})=\sigma^2_{\alpha} \where \sigma_{\alpha}>0\\
\eta_i &\indep  \alpha_i \indep \phi_i \indep \theta_i \indep \varepsilon_{i,t} \indep \t{\varepsilon}_{i}.
\end{align*} 
We further define 
$\tilde{\alpha}_i=\mu_{\alpha}+\delta_i'\mbf{x}_{i, T_i^*+1}$. 
We will investigate the post-shock aggregated estimators in $\mc{M}_2$ 
in settings where $\delta_i$ is either fixed or random. 
We let $\mc{M}_{21}$ denote model $\mc{M}_{2}$ with $\delta_i = \delta$ for $i= 1, \ldots, n+1$, 
where $\delta$ is a  fixed unknown parameter.
We let $\mc{M}_{22}$ denote model $\mc{M}_{2}$ with the following random effects 
structure for $\delta_i$:
\begin{align*}
\delta_i\simiid  \mc{F}_{\delta} \text{ with }\mrm{E}_{\mc{F}_{\delta}}(\delta_i)=\mu_{\delta}, \mrm{Var}_{\mc{F}_{\delta}}(\delta_i)=\Sigma_\delta 
   \quad \text{ with } \quad  \delta_i  \indep \t{\varepsilon}_{i}.
\end{align*}
We further define the parameter sets
\begin{align}
  \begin{array}{lll}
     \Theta &= &\{(\eta_i, \phi_i, \theta_i, \alpha_i, \mbf{x}_{i,t}, y_{i,t-1}, \delta_i)\colon t= 1, \ldots, T_i, i = 2, \ldots, n +1\},\\
    \Theta_1 &= &\{(\eta_i, \phi_i, \theta_i, \alpha_i, \mbf{x}_{i,t}, y_{i,t-1}, \delta_i)\colon t= 1, \ldots, T_i, i = 1\},\label{parameter}
  \end{array}
\end{align}
where $\Theta$ and $\Theta_1$ can adapt to $\mc{M}_1$ by dropping $\delta_i$. We assume this for notational simplicity.

\subsection{Forecast}
\label{forecast}
In this section we show how post-shock aggregate estimators improve upon standard 
forecasts that do not account for the shock effect.
%The interest of this study lies in comparing how consideration of shock effects 
%improves the prediction. 
More formally, we will consider the following candidate forecasts: 
\begin{align*}
  &\text{Forecast 1}: \hat y_{1,T_1^*+1}^1 = \hat\eta_1 
    + \hat\phi_1 y_{1,T_1^*} + \hat\theta_1'\x_{1,T_1^*+1} 
    , \\
  &\text{Forecast 2}: \hat y_{1,T_1^*+1}^2 = \hat\eta_1 
    + \hat\phi_1 y_{1,T_1^*} + \hat\theta_1'\x_{1,T_1^*+1} 
    + \hat{\alpha},
\end{align*}
where $\hat\eta_1$, $\hat\phi_1$, and $\hat\theta_1$ are all OLS estimators of $\eta_1$, $\phi_1$, and $\theta_1$, respectively, and $\hat{\alpha}$ is some form of estimator for the shock effect of time series of interest, i.e., $\alpha_1$. 
The first forecast ignores the presence of $\alpha_1$ while the second forecast 
incorporates an estimate of $\alpha_1$ that is obtained from the other independent forecasts under study. 
%Under $\mc{M}_1$ (Section \ref{modelsetup}), $\E{\alpha_1} = \mu_{\alpha}$.

Note that the two forecasts do not differ in their predictions for 
$y_{1,t}$, $t = 1,\ldots T_1^*$. Instead, they only differ in predicting 
$y_{1,T_1^*+1}$. Throughout the rest of this article we show that the donor pool of time series $\{y_{i,t} \colon t = 1,\ldots,T_i, i = 2,\ldots,n+1\}$ has 
the potential to improve the forecasts for $y_{1, T_1^*+1}$ under  different 
circumstances for the dynamic panel model $\mc{M}_1$, $\mc{M}_{21}$, and $\mc{M}_{22}$. Improvement will be measured by assessing the reduction in risk that Forecast 2 offers over Forecast 1. We will return to the theoretical details of risk-reduction in Section \ref{properties}.

We specifically focus on predictions for $y_{1,T_1^*+1}$, the first post-shock response. It is important to note that in general $\hat{\alpha}$ does not converge to $\alpha_1$ in any sense.  Despite this shortcoming, adjustment of the forecast for $y_{1,T_1^*+1}$ through the addition of $\hat{\alpha}$ has the potential to lower forecast risk under several conditions corresponding to different estimators of $\alpha_1$. % which will be discussed shortly.

\subsection{Construction of shock effects estimators}
\label{constructionofestimators}

We now construct the aggregated estimators of the shock effects that appear in Forecast 2 (see Section \ref{forecast}). We use these to forecast response values $y_{1, T_1^*+1}$ assuming that $T_1^*$ is known.  First, we introduce the procedures of parameter estimation for $\mc{M}_1$, $\mc{M}_{21}$, and $\mc{M}_{22}$ (see Section \ref{modelsetup}). For $i = 2, \ldots, n+1$, all parameters in this model will be estimated with ordinary least squares (OLS) using historical data of $t = 1, \ldots, T_i$. For $i = 1$, we estimate all the parameters but $\alpha_1$ using OLS procedures for $t=1, \ldots, T_1^*$. In particular, let $\hat{\alpha}_i$, $i = 2, \ldots, n+1$ be the OLS estimate of $\alpha_i$.  Note that parameter estimation for $\mc{M}_1$ is identically the same as that for $\mc{M}_{21}$ or $\mc{M}_{22}$.  We emphasize that $\alpha_i$s are random variables, but the OLS estimation is conditioned on the realizations from some distribution.

Second, we introduce the candidate estimators for $\alpha_1$. Define the 
adjustment estimator for time series $i=1$ by
\begin{equation} \label{adjusted}
  \hat{\alpha}_{\text{adj}} = \frac{1}{n}\sum_{i=2}^{n+1}\hat{\alpha}_i,
\end{equation}
where the $\hat{\alpha}_i$s in \eqref{adjusted} are OLS estimators of  
the $\alpha_i$s for $i = 2, \ldots, n+1$.  We can use $\hat{\alpha}_{\text{adj}}$ as an estimator for 
the unknown $\alpha_1$ term for which no meaningful estimation information 
otherwise exists. It is intuitive that $\hat{\alpha}_{\rm adj}$ should perform 
well under $\mc{M}_1$ where we assume that $\alpha_i$'s share the same mean 
for $i= 1, \ldots, n+1$. However, it can also be shown that 
$\hat{\alpha}_{\rm adj}$ may be less favorable in $\mc{M}_{21}$ 
and $\mc{M}_{22}$, which will be discussed in detail in Section \ref{properties}. 

We also consider the inverse-variance weighted estimator 
in practical settings where the $T_i$'s and $T_i^*$'s vary greatly across $i=2, \ldots, n+1$. 
The inverse-variance weighted estimator is defined as 
\begin{align*}
  \hat{\alpha}_{\rm IVW} = \frac{\sum_{i=2}^{n+1} \hat{\alpha}_i / \hat{\sigma}_{i\alpha}^2}{\sum_{i=2}^{n+1} 1/\hat{\sigma}_{i\alpha}^2},
  \quad \text{ where } \quad  \hat{\sigma}_{i\alpha}^2 = \hat{\sigma}^2_i( \mathbf{U}_i'\mbf{U}_i)_{22}^{-1},
\end{align*}
where  $\hat{\alpha}_i$ is the OLS estimator of $\alpha_i$, 
$\hat{\sigma}_i$ is the residual standard error from OLS estimation, 
and $\mbf{U}_i$ is the design matrix for OLS with respect to time series 
for $i = 2, \ldots, n+1$. Note that since $\sigma$ is unknown, estimation 
is required and the numerator and denominator terms are dependent in general. 
%It is hard to evaluate its expectation and variance. 
%But it is clear 
%that $\hat{\alpha}_{\rm IVW}$ will generally not be unbiased. 
However, $\hat{\alpha}_{\rm IVW}$ can be a reasonable estimator in 
practical settings. %when the change 
%point is not symmetric and/or there are some series with small or large number 
%of time points recorded. We will then investigate the performance of this 
%estimator in simulation (Section \ref{simulation}).
We do not provide closed form expressions for $\E{\hat{\alpha}_{\rm IVW}}$ 
and $\var{\hat{\alpha}_{\rm IVW}}$ but empirical performance of 
$\hat{\alpha}_{\rm IVW}$ is assessed via Monte Carlo simulation 
(see Section \ref{simulation}).

We now motivate a weighted-adjustment estimator for model $\mc{M}_{21}$ 
and $\mc{M}_{22}$. Our weighted-adjustment estimator is inspired by the 
weighting techniques in synthetic control methodology (SCM) developed 
in \cite{abadie2010synthetic}. 
%we intend to construct a weighted adjustment estimator by using similar but 
%different methods. 
However, our weighted-adjustment estimator is not a causal estimator and 
our estimation premise is a reversal of that in SCM. 
%The case study of \citet{abadie2010synthetic} in essence observes the data and 
%estimates the effect of policy. 
Our objective is in predicting a post-shock response $y_{1,T_1^*+1}$ that is not yet 
observed using other time series whose post-shock responses are observed.
%However, the idea of merging information from  similar events (i.e., time series in our 
%study) to improve the solutions for the problem of interest should be the same. 

We use similar notation as that in \cite{abadie2010synthetic} to motivate our weighted-adjustment estimator. Consider a $\mbf{W} \in \R^n$ weight vector 
$\mbf{W}=(w_2, \ldots, w_{n+1})'$, where $w_i\in [0,1]$ for all 
$i = 2, \ldots, n+1$. Construct
\begin{align*}
 \mbf{X}_1 = \mbf{x}_{1, T_1^*+1}',
  \quad  
  \mbf{X} = \begin{pmatrix}
    \mbf{x}_{2, T_2^*+1}' \\
    \vdots \\
    \mbf{x}_{n+1, T_{n+1}^*+1}'
  \end{pmatrix},
  \quad \text{and} \quad 
  \hat{\mbf{X}}_1(\mbf{W}) 
    = \mbf{W}'\mbf{X},
\end{align*}
where $\mbf{X}_1,\hat{\mbf{X}}_1(\mbf{W}) \in \R^{1 \times p}$. Define $\mc{W}=\{\mbf{W}\in [0,1]^n \colon 1_n'\mbf{W} = 1 \}$. 
Suppose there exists $\mbf{W}^*\in \mc{W}$ with 
$\mbf{W}^*=(w_2^*, \ldots, w_{n+1}^*)'$ such that
\begin{align}
 \mbf{X}_1=\hat{\mbf{X}}_1(\mbf{W}^*),  \quad i.e., \quad \mbf{x}_{1, T_1^*+1} = \sum_{i=2}^{n+1} w_i^*\mbf{x}_{i, T_i^*+1}.\label{SCM}
\end{align}
Note that (\ref{SCM}) tries to find $\mbf{W}^*$ such that $\mbf{x}_{1, T_1^*+1}$ is a convex combination of $\mbf{x}_{i, T_i^*+1}$ for $i = 2, \ldots, n+1$ with weights $\mbf{W}^*$. Therefore, $\mbf{W}^*$ should exist as long as $\mbf{X}_1$ falls in the convex hull of 
 \begin{align*}
   \left\{ \mbf{x}_{2, T_2^*+1}', \ldots, \mbf{x}_{n+1, T_{n+1}^*+1}' \right\}.
 \end{align*}
%It is a reasonable assumption that the pool of time series that we are considering 
%should be similar to the time series of interest. 
Our weighted-adjustment estimator will therefore perform well when the pool of time series posses similar covariates to the time series for which no post-shock responses are observed. We compute $\mbf{W}^*$ as
\begin{align}
  \mbf{W}^* = \argmin_{\mbf{W}\in \mc{W}} \norm{\mbf{X}_1-\hat{\mbf{X}}_1(\mbf{W})}_{p}. 
  \label{W}
\end{align}
%which is a more general form of $\mbf{W}^*$ for the case when $\mbf{X}_1$ does not fall in the convex hull of (\ref{convexhull}). 
\cite{abadie2010synthetic} commented that we can select $\mbf{W}^*$ 
so that (\ref{SCM}) holds approximately %so that the property of $\mbf{W}^*$ should 
%still hold approximately 
and that weighted-adjustment estimation techniques of this form are not 
appropriate when the fit is poor. 
Note that $\mbf{W}^*$ is not random since the covariates are assumed to be fixed. Since $\mc{W}$ is a closed and bounded subset of $\reals^n$,  $\mc{W}$ is compact. Because the objective function 
is continuous in $\mbf{W}$, $\mbf{W}^*$ will always exist. %Relying on this weight, 
%we are mainly interested in constructing the following adjustment estimator:
Our weighted-adjustment estimator for the shock effect $\alpha_1$ is
  \begin{align*}
    \hat{\alpha}_{\rm wadj} = \sum_{i=2}^{n+1} w_i^*\hat{\alpha}_i
    \quad \text{ for } \quad \mbf{W}^* = \begin{pmatrix}
      w^*_2 & \cdots & w^*_{n+1}
    \end{pmatrix}'.
  \end{align*}
 %Different from \citet{abadie2010synthetic}, our construction customizes synthetic control methods to the setting of AR(1) model. To be more specific, notice that (\ref{SCM}) implicitly uses the covariates at $T_*$ the shock-time and $T_*-1$ the time point just before the shock. This model somehow accounts for the impact of past information on the shock effects. 
 We further define
\begin{align*}
  \mathbf{V} = (\mathbf{x}_{2, T_2^*+1}, \ldots,\mathbf{x}_{n+1, T_{n+1}^*+1}).
\end{align*}
\begin{prop}
  \label{uniqueness} If $\mathbf{V}$ has full rank and it exists some $\mathbf{W}$ satisfies (\ref{SCM}), the solution to  (\ref{W}) is unique.
\end{prop} 
Proposition \ref{uniqueness} details some conditions when $\mathbf{W}^*$ is unique.  Note that $\mathbf{V}$ is $p \times n$. Therefore, if the covariates are of full rank and the true solution lies in the convex and compact $\mathcal{W}$, a sufficient condition for $\mathbf{W}^*$ to be unique is $p \geq n$. However, when $p < n$, $\mathbf{W}^*$ may not be unique. If it exists some $\mathbf{W}^*$ satisfies (\ref{SCM}) and $p < n$, there are infinitely many solutions to (\ref{SCM}).  The issue of non-uniqueness is further discussed in Section \ref{varbootstrap}.

\begin{remark}
% Though in Section \ref{modelsetup} we assume $\mbf{x}_{i,t}\in \reals^p$ and bases construction for $\hat{\alpha}_{\rm adj}$ and $\hat{\alpha}_{\rm wadj}$  on this fact, we will shortly show that the covariates need not be of the same dimension across disparate time series. 
\label{remark1}In Section \ref{modelsetup} we specify that $\mbf{x}_{i,t}, \theta_i \in \reals^p$. 
 % We can insert $\mathbf{0}$ into the covariates of disparate time series for the parts that they do not share. For example, suppose $\mathbf{x}_{2, t}, \mathbf{x}_{2, t-1}\in \reals^{p-1}$ and $\mathbf{x}_{3, t}, \mathbf{x}_{3, t-1}\in \reals^p$ are the covariates for time series 2 and 3. Assume that the $p$th column of $\mathbf{x}_{3, t}, \mathbf{x}_{3, t-1}$ is the part they do not share. In this case, we can let $\mathbf{x}^{adj}_{2, t}=(\mathbf{x}_{2, t}, \mathbf{0})$, $\mathbf{x}^{adj}_{2, t-1}=(\mathbf{x}_{2, t-1}, \mathbf{0})$ to satisfy the $p$-dimension requirement. 
However, it is not necessary that the all $p$ covariates are important for every time series under study. The regression coefficients $\theta_i$ are nuisance parameters that are not of primary importance. 
% In this regard, OLS estimations in software can reparameterize the design matrix.  Of course, the column space of the design matrix does not change compared to the one in original dimension. Thus, the least squares estimators do not change. In other words, OLS estimation need no adjustment. The generalization may only apply to the construction of weighted-adjustment  estimator $\hat{\alpha}_{\rm wadj}$.
It will be understood that structural 0s in $\theta_i$ correspond to variables that are unimportant. 
\end{remark}

\begin{remark}
  Our forecasting premise and estimation construction shares similarities with Bayesian viewpoints. From a Bayesian perspective, if we assign a prior $\pi$ to $\alpha_1$, $\hat{\alpha}_{\rm adj}$, $\hat{\alpha}_{\rm wadj}$, and $\hat{\alpha}_{\rm IVW}$ can be interpreted as the Bayes rules with respect to $\pi$ under different loss functions. If the sampling distribution of the data and $\pi$ are known, it is possible to compute the Bayes risks of $\hat{\alpha}_{\rm adj}$, $\hat{\alpha}_{\rm wadj}$, and $\hat{\alpha}_{\rm IVW}$ with respect to $\pi$,  thus enabling comparisons among them. Additionally, from Theorem 2.4 in Chapter 5 of \cite{lehmann2006theory}, $\hat{\alpha}_{\rm adj}$, $\hat{\alpha}_{\rm wadj}$, and $\hat{\alpha}_{\rm IVW}$ are admissible if they are unique with probability one.
\end{remark}

\section{Forecast risk and properties of shock effects estimators}
\label{properties}

In this section, we discuss the properties that are related to forecast-risk reduction. In discussion of risk, it is useful to derive expressions for expectation and variance of the adjustment estimator $\hat{\alpha}_{\rm adj}$ and weighted-adjustment estimator.  The expressions for the expectations are as follow,

 \begin{enumerate}[label = (\roman*)]
    \item Under $\mc{M}_{1}$, $\E{\hat{\alpha}_{\rm adj}}=\E{\hat{\alpha}_{\rm wadj}} = \mu_{\alpha}$.
    \item Under $\mc{M}_{21}$, 
    $
      \E{\hat{\alpha}_{\rm adj}} = \mu_{\alpha} + \frac{1}{n} \sum_{i=2}^{n+1} \delta' \mbf{x}_{i, T_i^*+1}
    $
    and
    $
       \E{\hat{\alpha}_{\rm wadj}} = \mu_{\alpha} + \delta'\mbf{x}_{1, T_1^*+1} .
    $
    \item Under $\mc{M}_{22}$,
    $
      \E{\hat{\alpha}_{\rm adj}} = \mu_{\alpha} + \frac{1}{n} \sum_{i=2}^{n+1} \mu_{\delta}' \mbf{x}_{i, T_i^*+1}
    $
    and
    $
       \E{\hat{\alpha}_{\rm wadj}} = \mu_{\alpha} + \mu_{\delta}'\mbf{x}_{1, T_1^*+1} .
    $
  \end{enumerate}
Formal justification for these results can be found in Appendix. Note that $\hat{\alpha}_{\rm adj}$, $\hat{\alpha}_{\rm wadj}$, and $\hat{\alpha}_{\rm IVW}$ are not unbiased estimators for $\alpha_1$. However, under $\mc{M}_{1}$, $\hat{\alpha}_{\rm adj}$ and $\hat{\alpha}_{\rm adj}$ are unbiased estimators for $\E{\alpha_1}=\mu_{\alpha}$. Nevertheless, $\hat{\alpha}_{\rm adj}$ is a biased estimator for $\E{\alpha_1}$ but $\hat{\alpha}_{\rm wadj}$ is an unbiased estimator for $\E{\alpha_1}$ under both $\mc{M}_{21}$ and $\mc{M}_{22}$. We collect these results in the following proposition. 

\begin{prop}
\label{unbiased} 
\quad 
\begin{enumerate}[label = (\roman*)]
  \item Under $\mc{M}_1$, $\hat{\alpha}_{\rm adj}$ is an unbiased estimator of $\E{\alpha_1}$. Under $\mc{M}_{21}$ and $\mc{M}_{22}$, $\hat{\alpha}_{\rm adj}$ is a biased estimator of $\E{\alpha_1}$ in general.
  \item Suppose that $\mbf{W}^*$ satisfies (\ref{SCM}). Under $\mc{M}_{1}$, $\mc{M}_{21}$ and $\mc{M}_{22}$, $\hat{\alpha}_{\rm wadj}$ is an unbiased estimator of $\E{\alpha_1}$.
\end{enumerate}
\end{prop}

Unbiasedness properties for $\E{\alpha_1}$ of $\hat{\alpha}_{\rm adj}$ and $\hat{\alpha}_{\rm wadj}$ allow for simple conditions for risk-reduction to hold, and more importantly motivates a bootstrap estimation for evaluation of these conditions. % for $\hat{\alpha}_{\rm adj}$ and  $\hat{\alpha}_{\rm wadj}$ to reduce risk, and make it more clear with respect to when one is better than the other. 
These conditions and bootstrap will be discussed in Section \ref{conditions} and \ref{varbootstrap}, respectively. Next, we present the variance expressions for $\hat{\alpha}_{\rm adj}$ and $\hat{\alpha}_{\rm wadj}$ as below.

\begin{enumerate}[label = (\roman*)]
  \item Under $\mc{M}_1$ and $\mc{M}_{21}$,  
\begin{align*}
  \var{\hat{\alpha}_{\rm adj}} 
  &=\frac{\sigma^2}{n^2}\sum_{i=2}^{n+1}\mrm{E}\big\{(\mbf{U}'_i\mbf{U}_i)^{-1}_{22}\big\}+\frac{\sigma^2_{\alpha}}{n^2}\\
\var{\hat{\alpha}_{\rm wadj}}  &= \sigma^2\sum_{i=2}^{n+1}(w_i^*)^2\mrm{E}\big\{(\mbf{U}'_i\mbf{U}_i)^{-1}_{22}\big\}+\sigma^2_{\alpha}\sum_{i=2}^{n+1}(w_i^*)^2
\end{align*}
\item Under $\mc{M}_{22}$, 
\begin{align*}
\var{\hat{\alpha}_{\rm adj}} 
  &=\frac{\sigma^2}{n^2}\sum_{i=2}^{n+1}\mrm{E}\big\{(\mbf{U}'_i\mbf{U}_i)^{-1}_{22}\big\}+\frac{1}{n^2}(\mbf{x}_{i, T_i^*+1}'\Sigma_{\delta}\mbf{x}_{i, T_i^*+1} + \sigma^2_{\alpha})\\
  \var{\hat{\alpha}_{\rm wadj}} 
  &= \sigma^2\sum_{i=2}^{n+1}(w_i^*)^2\mrm{E}\big\{(\mbf{U}'_i\mbf{U}_i)^{-1}_{22}\big\}
  + \sum_{i=2}^{n+1} (w_i^*)^2 (\mbf{x}_{i, T_i^*+1}'\Sigma_{\delta}\mbf{x}_{i, T_i^*+1} + \sigma^2_{\alpha}).
\end{align*}
\end{enumerate}
Formal justification for these results can be found in Appendix. Note that the variances are not comparable in closed-form %we shall see that it is difficult to compare variances between $\hat{\alpha}_{\rm adj}$ and $\hat{\alpha}_{\rm wadj}$ 
because of the term $\mrm{E}\big\{(\mbf{U}'_i\mbf{U}_i)^{-1}_{22}\big\}$.  This term exists because of the inclusion of the random lagged response in our autoregressive model formulation.

Section \ref{conditions} details conditions needed for risk-reduction and comparisons of adjustment estimators. These conditions involve variances and expectations which may be difficult to compute in practice. To make use of those conditions in practice, estimation is required. Sections \ref{varbootstrap} introduce a residual bootstrap procedure which estimates the involved parameters in those conditions and thus motivates prospective decision-making about whether $\hat{\alpha}_i$ reduces the risk. Section \ref{loocv} describes our leave-one-out cross validation procedures, which prospectively estimate the correctness of such decision without observation of the post-shock response for the time series under study. Our simulations verify these procedures.

\subsection{Risk-reduction conditions for shock effects estimators}
\label{conditions}

In this section we will discuss the conditions for risk reduction for individual shock effects estimators under $\mc{M}_1$, $\mc{M}_{21}$, and $\mc{M}_{22}$. %The reason is that if the estimator of $\alpha_1$ turns out to be an unbiased estimator of $\E{\alpha_1}$, the conditions get  simplified. 
For an adjustment estimator $\hat{\alpha}$, we will write the risk-reduction as $\Delta(\hat{\alpha})=R_{T_1^*+1,1}-R_{T_1^*+1,2}$ where $R_{T_1^*+1,2}$ is the risk of Forecast 2 calculated using the adjustment estimator $\hat{\alpha}$. 

\subsubsection{Conditions under $\mc{M}_1$}
 \label{conditionsmodel1}
 
Recall that Proposition \ref{unbiased} implies that the adjustment estimator $\hat{\alpha}_{\rm adj}$ and weighted-adjustment estimator $\hat{\alpha}_{\rm wadj}$ are unbiased for $\E{\alpha_1}$ under $\mc{M}_1$. With this result, we will have  the following propositions that specify the conditions that are necessary for risk reduction. 

\begin{prop}
\label{proprisk}Under $\mc{M}_1$,
\begin{enumerate}[label = (\roman*)]
  \item $\Delta(\hat\alpha_{\rm adj}) > 0$  when 
$\Var(\hat{\alpha}_{\rm adj}) < \mu_{\alpha}^2$.
  \item if $\mbf{W}^*$ satisfies (\ref{SCM}), then $\Delta(\hat\alpha_{\rm wadj}) > 0$ when $\var{\hat{\alpha}_{\rm wadj}}<\mu_{\alpha}^2$. 
\end{enumerate}
\end{prop}

Proposition \ref{proprisk} says that under $\mc{M}_1$ if the variance of the estimator is smaller than the squared mean of $\alpha_1$, those estimators will enjoy the risk reduction properties.   In this setting, under $\mc{M}_1$, $\Delta(\hat{\alpha}_{\rm adj})= \mu_{\alpha}^2-\var{\hat{\alpha}_{\rm adj}}$ and $\Delta(\hat{\alpha}_{\rm wadj})= \mu_{\alpha}^2-\var{\hat{\alpha}_{\rm wadj}}$. % from Proposition \ref{proprisk}.
From Proposition \ref{proprisk}, we obtain a risk-reduction condition
\begin{align}
  \var{\hat{\alpha}_{\rm adj}} 
  &=\frac{\sigma^2}{n^2}\sum_{i=2}^{n+1}\mrm{E}\big\{(\mbf{U}'_i\mbf{U}_i)^{-1}_{22}\big\}+\frac{\sigma^2_{\alpha}}{n^2} < \mu_{\alpha}^2. \label{riskconditionadj}
\end{align}

Condition (\ref{riskconditionadj}) implies two facts: (1) adjustment (Forecast 2) is preferable to no adjustment (Forecast 1) asymptotically in $n$ whenever $\mu_{\alpha} \neq 0$ (see Forecast in Section \ref{forecast}); (2) In finite donor pool settings, adjustment is preferable to no adjustment when $\mu_{\alpha}$ is large relative to its variability and overall regression variability.   %This result is also intuitive. For example, if  $\mu_{\alpha}$ is large, Forecast 1 will definitely work poor because large  $\mu_{\alpha}$ implies shock effects that cannot be ignored.  See Section \ref{forecast} for forecast formulation for details.

If  $\mathbf{W}^*$ does not satisfy (\ref{SCM}), its unbiased properties for $\E{\alpha_1}$ should hold approximately when the fit in (\ref{W}) is appropriate as commented in Section \ref{constructionofestimators}. From Proposition \ref{proprisk} and the variance expression for $\hat{\alpha}_{\rm wadj}$, the risk-reduction condition for $\hat{\alpha}_{\rm wadj}$ is
\begin{equation} \label{varwadj}
\var{\hat{\alpha}_{\rm wadj}}
 = \sigma^2\sum_{i=2}^{n+1}(w_i^*)^2\mrm{E}\big\{(\mbf{U}'_i\mbf{U}_i)^{-1}_{22}\big\}+\sigma^2_{\alpha}\sum_{i=2}^{n+1}(w_i^*)^2 < \mu_{\alpha}^2.
\end{equation}
In this case, adjustment is preferable to no adjustment when $\mu_{\alpha}$ is large relative to the weighted sum of variances for shock effects for other time series and overall regression variability.  However, the above criteria are generally difficult to evaluate in practice.  Sections \ref{varbootstrap} and \ref{loocv}  provide detailed treatments on how to estimate the sign of $\Delta(\hat\alpha)$ in practice.

\subsubsection{Conditions under $\mc{M}_{21}$ and $\mc{M}_{22}$}
\label{conditionsm2122}

The shock effects $\alpha_i$s have different means under $\mc{M}_{21}$ and $\mc{M}_{22}$ unlike under $\mc{M}_1$. %It is a more reasonable and general model since it is often the case that shock effects differ by means among disparate time series in practice. $\mc{M}_{22}$  further adds the random effect structure of $\delta$ to  $\mc{M}_{21}$ for generalization.
However, Proposition \ref{unbiased} implies that $\hat{\alpha}_{\rm wadj}$ is an unbiased estimator of $\E{\alpha_1}$. %From Proposition \ref{varprop}, the risk-reduction condition for $\hat{\alpha}_{\rm wadj}$  will be as in Proposition \ref{propriskwadj2}. 
We now state conditions for risk-reduction.

\begin{prop}
\label{propriskwadj2} If $\mbf{W}^*$ satisfies (\ref{SCM}), then $\Delta(\hat\alpha_{\rm wadj}) > 0$ when $\var{\hat{\alpha}_{\rm wadj}}<(\E{\alpha_1})^2$  under $\mc{M}_{21}$ and $\mc{M}_{22}$. 
\end{prop}
Under Proposition \ref{propriskwadj2}, we can obtain a risk-reduction inequality that is similar to \eqref{varwadj},
\begin{align*}
\var{\hat{\alpha}_{\rm wadj}}
 = \sigma^2\sum_{i=2}^{n+1}(w_i^*)^2\mrm{E}\big\{(\mbf{U}'_i\mbf{U}_i)^{-1}_{22}\big\} + \sum_{i=2}^{n+1} (w_i^*)^2 (\mbf{x}_{i, T_i^*+1}'\Sigma_{\delta}\mbf{x}_{i, T_i^*+1} + \sigma^2_{\alpha}) < (\E{\alpha_1})^2,
\end{align*}
where $\mbf{x}_{i, T_i^*+1}'\Sigma_{\delta}\mbf{x}_{i, T_i^*+1} + \sigma^2_{\alpha}$ may be replaced with $\sigma^2_{\alpha}$ in $\mc{M}_{21}$. The conclusions and intuitions will be identically the same as what we have in Section \ref{conditionsmodel1}. Proposition \ref{unbiased} shows that $\hat{\alpha}_{\rm adj}$ is a biased estimator of $\E{\alpha_1}$ under $\mc{M}_{21}$ and $\mc{M}_{22}$ generally. Hence, Proposition \ref{proprisk} no longer holds for $\hat{\alpha}_{\rm adj}$ under $\mc{M}_{21}$ and $\mc{M}_{22}$.

As an alternative, we can derive similar risk-reduction conditions that are appropriate for this setting. By Lemma \ref{risklemma} (see Section \ref{proofs}) and risk decomposition, we will achieve risk-reduction as long as
\begin{align*}
 \E{\alpha_1^2}= \var{\alpha_1}+(\E{\alpha_1})^2
 &>\E{\hat{\alpha}_{\rm adj}-\alpha_1}^2\\
  &=\var{\hat{\alpha}_{\rm adj}} +  (\E{\hat{\alpha}_{\rm adj}}-\alpha_1)^2 \\
  &=\var{\hat{\alpha}_{\rm adj}} +  \var{\alpha_1} + (\E{\hat{\alpha}_{\rm adj}}-\E{\alpha_1})^2.
\end{align*}
The above inequality simplifies to 
\begin{equation} \label{ineq}
 (\E{\alpha_1})^2 >\var{\hat{\alpha}_{\rm adj}}+ (\E{\hat{\alpha}_{\rm adj}}-\E{\alpha_1})^2.
\end{equation}

%A simple mean-squares decomposition produces the following inequality for assessing the favorability of $\hat{\alpha}_{\rm wadj}$ to $\hat{\alpha}_{\rm adj}$ under models $\mc{M}_{21}$ and $\mc{M}_{22}$,
% \begin{align*}
%  \var{\hat{\alpha}_{\rm adj}} 
%  -\var{\hat{\alpha}_{\rm wadj}} + \big(\E{\hat{\alpha}_{\rm adj}}-\mrm{E}(\alpha_1)\big)^2> 0.
%\end{align*}

%If it turns out to be fact that the variance of the weighted-adjustment estimator is greater than that of adjustment estimator, we should be aware that  the compromise for variance because of using $\hat{\alpha}_{\rm wadj}$ shouldn't exceed the squared bias, i.e., $\big(\E{\hat{\alpha}_{\rm adj}}-\mrm{E}(\alpha_1)\big)^2$.

As mentioned in Section \ref{constructionofestimators}, it is difficult to evaluate the expectation and variance of $\hat{\alpha}_{\rm IVW}$. We note that $\hat{\alpha}_{\rm IVW}$ is generally biased for $\E{\alpha_1}$. That is to say we can adapt the above proof to derive the risk-reduction conditions for $\hat{\alpha}_{\rm IVW}$: under $\mc{M}_{1}$, $\mc{M}_{21}$, and $\mc{M}_{22}$, $\Delta(\hat\alpha_{\rm IVW}) > 0$ when $\var{\hat{\alpha}_{\rm IVW}} +(\E{\hat{\alpha}_{\rm IVW}}-\E{\alpha_1})^2<(\E{\alpha_1})^2$.  In fact, more generally, using similar proof of Lemma \ref{risklemma}, it can be shown that under $\mc{M}_2$, the risk-reduction quantities are
\begin{align*}
  \Delta(\hat{\alpha}_{\rm adj}) 
  &= (\E{\alpha_1})^2 -\var{\hat{\alpha}_{\rm adj}}-(\E{\hat{\alpha}_{\rm adj}}-\E{\alpha_1})^2,\\
  \Delta(\hat{\alpha}_{\rm IVW}) 
  &= (\E{\alpha_1})^2 -\var{\hat{\alpha}_{\rm IVW}}-(\E{\hat{\alpha}_{\rm IVW}}-\E{\alpha_1})^2 ,\\
  \Delta(\hat{\alpha}_{\rm wadj}) 
  &= (\E{\alpha_1})^2 -\var{\hat{\alpha}_{\rm adj}},
\end{align*}
where we estimate $\Delta(\hat{\alpha})$ for estimator $\hat\alpha$  using bootstrap and leave-one-out cross validation procedures developed in Sections \ref{varbootstrap} and \ref{loocv}.

\subsection{Bootstrap for risk-reduction evaluation problems}
\label{varbootstrap}

In this section, we present bootstrap procedures that approximate the distribution of our shock effect estimators, checks the underlying conditions of our risk reduction propositions, and estimate risk-reduction quantity using plug-in approach in practice. Our procedure involves the resampling of residuals in the separate OLS fits. This procedure has its origins in Section 6 of \citet{efron1986bootstrap} and Chapter 12 of \cite{kilian2017structural}. Our procedure involves the resampling of the residuals which are assumed to be the realizations of an iid process.

Our first bootstrap procedure is as follows: Let $B$ be the bootstrap sample size. At iteration $b$, first resample the indices $I = \{2, \ldots, n+1\}$ of the donor pool with replacement to form $I^{(b)}$ with cardinality $n$, where we note that the elements of $I^{(b)}$ may not be unique in terms of their  indices in the donor pool. Initialize $y_{i,0}^{(b)}=y_{i,0}$ for all $i \in I^{(b)}$. Then, resample the residuals under models $\mc{M}_1$ or $\mc{M}_{2}$, compute the bootstrapped response  $y_{i,t}^{(b)}$ for   $t \in \{1, \ldots, T_i\}$ using the model estimated by original data,  and obtain shock effect estimators for each of the time series in the donor pool for all $i \in I^{(b)}$. These shock effect estimators are then used to construct any of the adjustment estimators $\hat{\alpha}^{(b)}_{\mrm{adj}}$, $\hat{\alpha}^{(b)}_{\mrm{wadj}}$, and $\hat{\alpha}^{(b)}_{\mrm{IVW}}$, for $b = 1,\ldots,B$. We can then estimate distributional quantities of our shock effect estimators under our considered models with the bootstrap samples $\hat{\alpha}^{(b)}_{\mrm{adj}}$, $\hat{\alpha}^{(b)}_{\mrm{wadj}}$, and $\hat{\alpha}^{(b)}_{\mrm{IVW}}$, for $b = 1,\ldots,B$. We denote this procedure by $\mc{B}_u$. We motivate a second bootstrap procedure $\mc{B}_f$ which treats the the donor pool as fixed, and not a realization from an infinite super-population. Therefore, there is no resampling of the donor pool in $\mc{B}_f$, it is otherwise similar to $\mc{B}_u$. An algorithmic formulation of $\mc{B}_u$ and $\mc{B}_f$  are outlined in Section 2 in the Supplementary Materials.

We will explicitly use these bootstrapped samples of shock effect estimators to check the risk-reduction conditions in Propositions \ref{proprisk} and \ref{propriskwadj2}. Recall that $\hat{\alpha}_{\rm adj}$,  $\hat{\alpha}_{\rm wadj}$ and $\hat{\alpha}_{\rm IVW}$ are unbiased estimators of their expectations, and $\hat{\alpha}_{\rm wadj}$ is an unbiased estimator of $\E{\alpha_1}$ under $\mc{M}_1$ and $\mc{M}_2$ from Proposition \ref{unbiased}. Our bootstrap procedure estimates the variance of our adjustment estimators. We can then estimate the risk-reduction propositions and inequalities. For example, we can estimate $\Delta(\hat{\alpha}_{\rm adj})$ under model $\mc{M}_{21}$ or $\mc{M}_{22}$ with 
\begin{align*}
%  \Delta(\hat{\alpha}_{\rm adj})
  %&= (\E{\alpha_1})^2 -\var{\hat{\alpha}_{\rm adj}} -(\E{\hat{\alpha}_{\rm adj}}-\E{\alpha_1})^2 \\
  \hat{\Delta}(\hat{\alpha}_{\rm adj}) & = (\hat{\alpha}_{\rm wadj})^2 -S^2_{\hat{\alpha}_{\rm adj}} -(\hat{\alpha}_{\rm adj}-\hat{\alpha}_{\rm wadj})^2,
\end{align*}
where $S^2_{\hat{\alpha}_{\rm adj}}$ is the bootstrap sample variance estimator for $\Var(\hat{\alpha}_{\rm adj})$. 

%Suppose we have a pool of shock effect estimators, $\mc{A}$, e.g., $\mc{A}= \{\hat{\alpha}_{\rm adj}, \hat{\alpha}_{\rm wadj}, \hat{\alpha}_{\rm IVW}\}$ in our study. We can determine which one is the best shock effect estimator. Suppose the $\hat{\Delta}(\hat{\alpha}_{\rm wadj})$ and $\hat{\Delta}(\hat{\alpha}_{\rm IVW})$ are estimators for $\Delta(\hat{\alpha}_{\rm wadj}) $ and $\Delta(\hat{\alpha}_{\rm IVW})$, respectively. Then, we can evaluate the risk-reduction propositions by judging whether $\hat{\Delta}(\hat{\alpha})>0$ for $\hat{\alpha}\in \mc{A}$. Besides, we can also select the best shock effect estimator accordingly. Define
%\begin{align}
%	\hat{\alpha}_{\rm best} = \argmax_{\hat{\alpha}\in \mc{A}} \hat{\Delta}(\hat{\alpha}) \quad \text{ and } \quad  \alpha_{\rm best} = \argmax_{\hat{\alpha}\in \mc{A}} \Delta(\hat{\alpha}). \label{best}
%\end{align}
%That is, the best estimator $\hat{\alpha}_{\rm best}$ is the one with the maximum estimated risk-reduction quantities whereas $\alpha_{\rm best}$ is the true best shock effect estimator given observed post-shock response. See more discussions in Section \ref{loocv}.

We reiterate the philosophical distinction between $\mc{B}_u$ and $\mc{B}_f$. $\mc{B}_u$ treats the donor pool as realizations from some infinite super-population of potential donors. In contrast, $\mc{B}_f$ treats the donor pool as being fixed  and known before the analysis is conducted, where the randomness arises from parameters and idiosyncratic error. A double bootstrap procedure with similar steps can estimate the distribution of $\hat\Delta(\hat{\alpha})$ for $\hat{\alpha} \in \mc{A}$. The double bootstrap, instead of checking whether $\Delta(\hat{\alpha})>0$, can check whether a bootstrap percentile interval of resampled estimates of $\Delta(\hat{\alpha})$ contain 0 at a desired error threshold. We investigated such a double bootstrap procedure and found that it produced inferences that were similar to those produced using the bootstrap techniques developed in the main text.

We stress that our bootstrap approximations cannot alleviate the inherent bias of using our adjustment estimators as surrogates for $\alpha_1$. We caution that the bootstrapping residuals in OLS estimation may not provide valid inference in moderate or high dimension where $p < T_i$ but $p / T_i$ is not close to zero for $i\in \{2, \ldots, n+1\}$ \citep{el2018can}; see alternatives for residual bootstrapping in linear models in \citet{el2018can}. 

Recall that $\mathbf{W}^*$  may not be unique if the conditions in Proposition \ref{uniqueness} are not satisfied. Non-uniqueness might be a concern theoretically. This is due to the fact that infinitely many different weights can lead to infinitely many non-unique $\hat{\alpha}_{\rm wadj}$'s all targeting on the same $\alpha_1$. For example, consider the case where the size of  donor pool to be 2, $\var{\hat{\alpha}_2}=1$, $\var{\hat{\alpha}_3}=2$, and there are two solutions to (\ref{SCM}), say, $\mathbf{W}^*_1=(1,0)$ and  $\mathbf{W}^*_2=(0,1)$. In this scenario, the weighted adjustment estimator induced by $\mathbf{W}^*_1$ has variance 1 whereas the one by $\mathbf{W}^*_2$ has variance 2. Nevertheless, even if $\hat{\alpha}_i$ has the same variance across $i = 2, \ldots, n+1$, the same issue would occur if there were infinitely many $\mbf{W}^*$ with different norms. It is possible to resolve this issue by selecting a unique weight $\mbf{W}^*$ that optimizes a desirable objective function, prior to which one should find the bases spanning the subspace of $\mbf{W}^*$ satisfying (\ref{SCM}). Simulations in the Section 4 of the Supplementary Materials provide some evidence that non-uniqueness of $\mathbf{W}^*$ is not problematic for inferences.

\subsection{Leave-one-out cross validation}
\label{loocv}

In this section, we adapt leave-one-out cross validation (LOOCV) to our estimation context in order to provide prospective evaluations of our adjustment techniques. Our proposed LOOCV procedure has its roots in Section 7.10 of \citet{hastie2009elements}.  Recall in Section \ref{modelsetup} that we are given the data $\{(\mbf{x}_{i,t}, y_{i,t}) \colon i = 1, \ldots, n+1, t = 1, \ldots, T_i\}$, where $\{(\mbf{x}_{1,t}, y_{1,t})\colon t = 1, \ldots, T_1\}$ is the data of the time series of interest and the remaining observations form the donor pool. For iteration $m \in \{1,\ldots,n\}$ of our LOOCV procedure, we set aside $\{(\mbf{x}_{m + 1, t}, y_{m + 1, t}) \colon t = 1, \ldots, T_{m+1}\}$ as the time series of interest, and construct a new donor pool $\{(\mbf{x}_{i, t}, y_{i, t}) \colon i \in \mc{I}_m, t = 1, \ldots, T_{i}\}$, where $\mc{I}_m=\{2, \ldots, n+1\} \setminus \{m+1\}$. Since the post-shock response $y_{m+1, T_{m+1}^*+1}$ is observed, we can evaluate the performance of our adjustment estimators and the original forecast made without adjustment (i.e., Forecast 1 in Section \ref{forecast}).

LOOCV can be very computationally intensive when $n$ is large, especially when combined with bootstrapping. To alleviate these concerns we can perform LOOCV with a random subset of $k \leq n$ iterations selected without replacement. In this setting, we let $\mc{J}$ be the randomly sampled indices. For $m \in \mc{J}$, we set aside $\{(\mbf{x}_{m + 1, t}, y_{m + 1, t}) \colon t = 1, \ldots, T_{m+1}\}$ as the time series of interest, and construct a new donor pool $\{(\mbf{x}_{i, t}, y_{i, t}) \colon i \in \mc{I}, t = 1, \ldots, T_{i}\}$, where $\mc{I}=\{2, \ldots, n+1\} \setminus \{m+1\}$. Based on the new donor pool, we estimate relevant parameters using bootstrap procedures outlined in Section \ref{varbootstrap}. In other words, $k$ times of bootstrapping are nested in a LOOCV procedure.  We find that $k=5$ or $k=10$ iterations of LOOCV performs well.

We now outline how LOOCV can be used to prospectively assess the performance of adjustment estimators. Let $\mc{A}$ be the set of adjustment estimators. For each $\hat{\alpha} \in \mc{A}$, let $\delta_{\hat{\alpha}} = I(\hat\Delta(\hat{\alpha})>0)$ be a decision rule where $I(\cdot)$ is the indicator function and a $1$ corresponds to the decision to use  estimator $\hat\alpha$. If $\Delta(\hat{\alpha})>0$ ($\Delta(\hat{\alpha})<0$, respectively) but $\delta_{\hat{\alpha}}$ incorrectly reported 1 (0, respectively) so that it makes the decision not to use $\hat{\alpha}$ (to use $\hat{\alpha}$, respectively), $\delta_{\hat{\alpha}}$ is said to be incorrect. If $\Delta(\hat{\alpha})<0$ ($\Delta(\hat{\alpha})>0$, respectively) and  $\delta_{\hat{\alpha}}$ correctly reported 0 (1, respectively) so that it makes the decision to use $\hat{\alpha}$, $\delta_{\hat{\alpha}}$ is said to be correct. These situations are depicted in the following table: \vspace*{0.3cm}

\begin{center}
  \begin{center}
      \begin{tabular}{cc|c|c}
        \hline
        & & \multicolumn{2}{c}{Decision} \\
        & & $\delta_{\hat{\alpha}} = 1$ & $\delta_{\hat{\alpha}} = 0$ \\ 
                \hline
     \multirow{2}{*}{Truth}  & $\Delta(\hat{\alpha})>0$ & Correct & Incorrect \\
      \cline{3-4}
      & $\Delta(\hat{\alpha})<0$  & Incorrect & Correct \\
      \hline
      \end{tabular}
  \end{center}
\end{center}
%Similar definitions can be adapted to the best shock effect estimators. 
\vspace*{0.3cm}

We will use 
$
	\mc{C}(\delta_{\hat\alpha}) 
	= I(\delta_{\hat{\alpha}} \text{ is correct})
 %\quad \text{ and } \quad  	
 %\mc{C}(\mc{A}) 
%	= I(\hat{\alpha}_{\rm best} = \alpha_{\rm best})
$
as a metric that evaluates the performance of forecasts made with the adjustment estimator $\hat\alpha$. If $\E{\mc{C}(\delta_{\hat{\alpha}})} > 0.5$, we claim that $\delta_{\hat{\alpha}}$ is better than random guessing. Note that $\mc{C}(\delta_{\hat{\alpha}})$ can generally be computed only when the post-shock response is observed. However, it is possible to estimate $\E{\mc{C}(\delta_{\hat{\alpha}})}$ using LOOCV. %For each $\hat{\alpha}\in \mc{A}$ we can compute the consistency as $\mc{C}^{(-m)}(\delta_{\hat{\alpha}})$ using the estimation procedures in Section \ref{varbootstrap}.
The LOOCV estimates for $\E{\mc{C}(\delta_{\hat{\alpha}})}$ are
\begin{equation} \label{loocvm}
	 \bar{\mc{C}}(\delta_{\hat{\alpha}})= \frac{1}{n} \sum_{m = 1}^n \mc{C}^{(-m)}(\delta_{\hat{\alpha}}),
	\end{equation}
where $\mc{C}^{(-m)}(\delta_{\hat{\alpha}})$ is computed with respect to donor pool with index set $\mc{I}_m$ and the $m+1$ time series is treated as the time series of interest.  The LOOCV with $k$ random draws estimates $\E{\mc{C}(\delta_{\hat{\alpha}})}$ as
\begin{equation} \label{loocvk}
	 \bar{\mc{C}}^{(k)}(\delta_{\hat{\alpha}})= \frac{1}{k} \sum_{m \in \mc{J}} \mc{C} ^{(-m)}(\delta_{\hat{\alpha}})	,
\end{equation}
where $\mc{J}$ is the set of the $k$ randomly sampled indicies. %Similarly, assuming the candidates in the donor pool are  mutually independent and the data  satisfy $\mc{M}_{2}$, $\bar{\mc{C}}^{(k)}(\delta_{\hat{\alpha}})$ and $\bar{\mc{C}}^{(k)}(\mc{A})$ should be almost unbiased estimates of $\E{\mc{C}(\delta_{\hat{\alpha}})}$ and $\E{\mc{C}(\mc{A})}$, respectively. The estimation bias should decrease as $k \to n$.

\begin{remark}
  Note that we allow the time series within the donor pool to be dependent but donor pool should be independent of the time series of interest. However, if we assume the mutual  independence structure, $\bar{\mc{C}}(\delta_{\hat{\alpha}})$ will be an almost unbiased estimator of $\E{\mc{C}(\delta_{\hat{\alpha}})}$ \citep[Page 222]{msos}. %In other words, mutual independence assumption is not related to  $\E{\mc{C}(\delta_{\hat{\alpha}})}$ (i.e., the intrinsic correctness) but its estimation. 
\end{remark}

\section{Numerical Examples}
\label{simulation}

\subsection{Modeling setup}

In this section we provide justification for our methods based on Monte Carlo simulation. We implemented our simulation based on $\mc{M}_{22}$ with negligibly small $\Sigma_{\delta}$ approximating the design of $\mathcal{M}_{21}$.  We consider $p=25$ and $\mu_{\alpha}=2$, where $p = 25$ is set to satisfy conditions in Proposition \ref{uniqueness}. Parameter setup of our simulations is detailed as follows: the $\phi_i$'s are sampled independently from $\mrm{Uniform}(0,1)$. We sampled $T_i$'s independently from  $\text{Gamma}(15, 10)$ that are further rounded to integers, where the minimum allowable value of $T_i$ is fixed to be 90. We will randomly draw $T_i^*$ from $\{p + 4, \ldots, T_i-1\}$. The choices of $T_i$ and $T_i^*$ are set up to satisfy a necessary condition for the design matrix of OLS estimation to have full rank. Moreover, it is  designed to illustrate the performance of $\hat{\alpha}_{\rm IVW}$ that may perform well in time series with varying lengths. Additionally, we generated the covariates from $\text{Gamma}(1,2)$ to set up a setting when the $\hat{\alpha}_{\rm wadj}$ may perform well. Last, we set $\delta_i\simiid  \normal{1}{0.5}$ and $\theta_i \sim \normal{0}{1}$. We will consider parameter setup by varying $\sigma$ in the model of $y_{i,t}$, $n$, the donor pool size, and $\sigma_{\alpha}$ in the model of $\alpha_i$. We choose a Monte Carlo sample size of $30$ replications and a bootstrap sample size of $B = 200$ for computation. Means and standard errors for estimated quantities will be recorded. Our LOOCV procedure will consider $k=5$ random draws. Recall in Section \ref{loocv} that $B$ times of bootstrap are nested in a LOOCV with $k$ random draws. It implies that $B(k+1)$ times of bootstrap replications are required for each Monte Carlo simulation.

\subsection{Performance metrics}

Our adjustment estimators will be evaluated by multiple criteria. We interpret $\delta_{\hat{\alpha}}= I(\hat{\Delta}(\hat{\alpha})>0)$ for $\hat{\alpha}\in \mc{A}$ as the guess, with 1 indicating that $\hat{\alpha}$ provides risk-reduction over the simple no-adjustment forecast, and 0 indicates the converse. We will consider the LOOCV estimators \eqref{loocvm} and \eqref{loocvk} to assess correct decision making. We will also consider the Euclidean distance between the post-shock forecasts $\hat{y}_{1,T_1^*+1}$, $\hat{y}_{1,T_1^*+1}+ \hat{\alpha}_{\rm adj}$, $\hat{y}_{1,T_1^*+1}+\hat{\alpha}_{\rm wadj}$, and $\hat{y}_{1,T_1^*+1}+\hat{\alpha}_{\rm IVW}$ and the realized post-shock response $y_{1,T_1^*+1}$. The first two metrics can combine to assess our forecasting methodology prospectively while the latter requires the realization of the post-shock response $y_{1,T_1^*+1}$.

\subsection{Monte Carlo results}
\label{parametricbootstrapsimulation}

In this section, we discuss simulation results for the bootstrap procedures used in estimating parameters for risk-reduction propositions and inequalities. We mainly discuss simulations under $\mc{M}_2$ (see Section \ref{modelsetup}) for $\mc{B}_u$ and $\mc{B}_f$ (see Section \ref{varbootstrap}) with comparisons to those under $\mc{M}_1$ whose results are listed in Section 3 in the Supplementary Materials. Two simulation setups are investigated. 

In the first simulation setup, we consider the parameter combination of  $n \in \{5, 10, 15, 25\}$ and $\sigma_{\alpha} \in  \{5, 10, 25, 50, 100\}$ where we fix $\sigma=10$. Note that $\E{\E{\alpha_1}}=52$, where the last expectation is operated under the density of the covariates. In other words, data with $\sigma_{\alpha} \in  \{5, 10, 25, 50, 100\}$  should well represent the situations when the signal of the covariates is strong and when it is nearly lost. Results are displayed in Table~\ref{table1} in the Appendix, Section \ref{tablesappendix}.

In the second simulation setup, we consider the parameter combination of $\sigma, \sigma_{\alpha} \in  \{5, 10, 25, 50, 100\}$ where we fix $n=10$. Likewise, $\sigma, \sigma_{\alpha} \in  \{5, 10, 25, 50, 100\}$ will produce situations when the signal of the covariates is strong and when it is nearly lost in the model of  both $y_{i,t}$ and $\alpha_i$. Results are  in Table~\ref{table2} in the Appendix, Section \ref{tablesappendix}.

First, assuming that $\bar{C}^{(k)}(\delta_{\hat{\alpha}})$ well estimates $\E{\mc{C}(\delta_{\hat{\alpha}}})$ and fixing $n$, we  observe from Table \ref{table1} that the decision making of $\delta_{\hat{\alpha}}$ is nearly correct  for $\hat{\alpha}\in \mc{A}$ when $\sigma_{\alpha}$ is small  from Table \ref{table1}. The reasons can be explained as follows. When $\sigma_{\alpha}$ is small, the signal of the covariates is strong so that $\hat{\alpha}_{\rm wadj}$ will be expected to capture the signal according to construction of $\hat{\alpha}_{\rm wadj}$ in Section \ref{constructionofestimators}. Moreover, when $\sigma_{\alpha}$ is small, $\mc{M}_{22}$ approximates $\mc{M}_{21}$ such that estimation of $\E{\alpha_1}$ should be nearly unbiased according to Proposition \ref{unbiased}. However, when the signal of the covariates is poor ($\sigma_{\alpha}$ is large), the decision rule $\delta_{\hat{\alpha}}$  becomes unreliable for $\hat{\alpha}\in \mc{A}$. It is to be expected since the bootstrap estimates become more biased. However, users can be warned by $\bar{C}^{(k)}(\delta_{\hat{\alpha}})$ to have an idea of the effectiveness of $\delta_{\hat{\alpha}}$. Second, fixing $\sigma_{\alpha}$, we can observe that the correctness of $\delta_{\hat{\alpha}}$ increases when $n$ increases. It is due to the robustness gain in estimation when $n$ increases.

Additionally, we  observe that in most cases $\delta_{\hat{\alpha}_{\rm wadj}}$ reports $\hat{\alpha}_{\rm wadj}$ reduces the risk even when $\bar{\mc{C}}^{(k)}(\delta_{\hat{\alpha}_{\rm wadj}})$ starts to break down, though they  follow similar patterns.   Recall from Section \ref{varbootstrap} that $\hat{\Delta}(\hat{\alpha})$ contains the squared bias for estimating $\E{\alpha_1}$. But it is not present for $\hat{\Delta}(\hat{\alpha}_{\rm wadj})$ since we applied the fact  $\hat{\alpha}_{\rm wadj}$ is unbiased for $\E{\alpha_1}$ from Proposition \ref{unbiased} in plugging it in with replacing $\E{\alpha_1}$. Therefore, when the signal from covariates is poorer, $\delta_{\hat{\alpha}_{\rm wadj}}$ becomes less conservative. Besides, the averaged $I\big(\hat{\Delta}(\hat{\alpha})>0\big)$ times $\bar{\mc{C}}^{(k)}(\delta_{\hat{\alpha}})$ can provide an approximation for the probability that $\hat{\alpha}$ actually reduces the risk assuming an symmetry of correctness between the cases when $\hat{\Delta}(\hat{\alpha})>0$ and when $\hat{\Delta}(\hat{\alpha})<0$. For example, when $n = 5$ and $\sigma_{\alpha}=50$, the probability that $\hat{\alpha}_{\rm adj}$ reduces the risk is approximately $0.83 \times 0.59 = 0.490$ from Table \ref{table1}.  In other words,  the probability that $\hat{\alpha}$ reduces the risk has the same pattern as $\bar{\mc{C}}^{(k)}(\delta_{\hat{\alpha}})$ has with $n$ and $\sigma_{\alpha}$ for $\hat{\alpha}\in \mc{A}$.

From columns related to distance to $y_{1, T_1^*+1}$ in Table \ref{table1}, as $\sigma_{\alpha}$ increases, the prediction appears to be poorer. When $\sigma_{\alpha}= 5, 10, 25$, forecasts using $\hat{\alpha}_{\rm adj}$, $\hat{\alpha}_{\rm wadj}$, and $\hat{\alpha}_{\rm IVW}$ are always better than the original forecast significantly. But it does not hold generally for the case when $\sigma_{\alpha}=50, 100$. It is reasonable in that when the $\sigma_{\alpha}$ is large, it is difficult to find a reliable estimate of $\alpha_1$.  Nevertheless, no statistical evidence has been found to support the claim that $n$ matters in prediction. In other words, the size of the donor pool matters for producing reliable decision-making of $\delta_{\hat{\alpha}}$ rather than reliable prediction.

From Table \ref{table2}, we observe that as $\sigma_{\alpha}$ increases fixing $\sigma$, $\bar{\mc{C}}(\delta_{\hat{\alpha}})$   decreases, which is a pattern similar to the one shown in the first experiment. Furthermore, as $\sigma$ increases fixing $\sigma_{\alpha}$, $\bar{\mc{C}}(\delta_{\hat{\alpha}})$ decreases as well. Note that the correctness  hinges on the estimation of the parameters. Since $\hat{\alpha}_{\rm wadj}$  is a linear combination of OLS estimates, as $\sigma$ increases, $\var{\hat{\alpha}_{\rm wadj}}$ increases as well. Therefore, $\hat{\alpha}_{\rm wadj}$ become more volatile and its estimation of $\E{\alpha_1}$ can be less reliable. Those reasons can explain  why an increase of $\sigma_{\alpha}$ contributes to a decrease of  $\bar{\mc{C}}(\delta_{\hat{\alpha}})$.  We observe similar patterns for distance to $y_{1, T_1^*+1}$ as well. When $\sigma$ increases with fixing $\sigma_{\alpha}$, it is likely that the degree of variation of $y_{1,t}$ exceeds the extent of adjustment improvement $\hat{\alpha}$ can contribute to for $\hat{\alpha}\in \mc{A}$.

With respect to averaged $I(\hat{\Delta}(\hat{\alpha})>0)$ (i.e., the guess), it starts to decrease as $\sigma$ increases. This is reasonable if we believe the bootstrap estimate $S^2_{\hat{\alpha}}$ provides a good approximation for $\var{\hat{\alpha}}$ for $\hat{\alpha}\in \mc{A}$. The reasons are outlined as follows: Recall in Section \ref{conditionsm2122}, the conditions of risk-reduction propositions involve $(\E{\alpha_1})^2 > \var{\hat{\alpha}} + (\E{\hat{\alpha}}-\E{\alpha_1})^2$ for $\hat{\alpha}\in \mc{A}$. Notice that $\var{\hat{\alpha}}$ is an increasing function of $\sigma$ since $\hat{\alpha}$ is estimated by OLS. Therefore, it explains the reason why the increase of $\sigma$ would result in a decrease of averaged $I(\hat{\Delta}(\hat{\alpha})>0)$ since the inequality is not likely to hold when  $\var{\hat{\alpha}}$ increases.

Simulation for $\mc{B}_f$ with the same parameter setup as that of $\mc{B}_u$ are implemented. See Table \ref{table3} and Table \ref{table4} for results in the Appendix, Section \ref{tablesappendix}. Comparing Table \ref{table1}
 and Table \ref{table2} yields that when $n$ is moderately small ($n = 10$) and $\sigma_{\alpha}$ is small ($\sigma_{\alpha}=5$), $\mc{B}_u$ is better than $\mc{B}_f$ with statistical evidence. For other situations, $\mc{B}_u$  and $\mc{B}_f$  are rather similar. It is likely that the extra randomness from sampling with replacement from donor pool compensates for the possible noises from a small donor pool. Concerning  Table \ref{table2} and Table \ref{table4}, it appears that when $n = 10$ and $\sigma_{\alpha}=5$, $\mc{B}_u$ is better than  $\mc{B}_f$  when $\sigma$ increases. It might be the case that additional layer of bootstrap in the donor pool buffers the negative effects on $\bar{\mc{C}}(\delta_{\hat{\alpha}})$  introduced from increasing variation of $y_{i,t}$. However, when $\sigma_{\alpha}$ increases over 5 and $n = 10$,  $\mc{B}_f$ and $\mc{B}_u$ are quite similar under  situations of different $\sigma$ and $\sigma_{\alpha}$. In conclusion, $\mc{B}_u$ is better than $\mc{B}_f$ when the signal of the covariates is strong and $n$ is moderately small; otherwise, they are similar. %See more discussions for differences between $\mc{B}_u$ and $\mc{B}_f$ in Section \ref{discussion}. 

 Simulation results corresponding to model $\mc{M}_1$ are listed in Section 3 in Supplementary Materials. Results under model $\mc{M}_1$ are very similar to those of $\mc{M}_2$, except for the difference among estimators. The results show that (1) the performance of $\hat{\alpha}_{\rm adj}$ and  $\hat{\alpha}_{\rm IVW}$ are nearly the same and (2) in many situations, $\hat{\alpha}_{\rm adj}$ and  $\hat{\alpha}_{\rm IVW}$ are  better than $\hat{\alpha}_{\rm wadj}$; in other situations, they are mostly the same. Recall that in $\mc{M}_1$, the models for $\alpha_1$ do not involve the covariates. Therefore, similarity weighting may not be informative when the model for $\alpha_i$ is identified wrongly. Under $\mc{M}_1$, simple averaging, aimed for a reduction of variance, or inverse-variance weighting, targeting on reducing negative effects from varying time lengths, may work better.

We have implicitly assumed that $\mathbf{W}^*$ is non-degenerate in the population in these simulations.  Recall that in Section \ref{constructionofestimators} we noted that if there exists some $\mathbf{W}^*$ which satisfies \eqref{SCM} and $p < n$, then there will be infinitely many solutions to $\mathbf{W}^*$. In this scenario, $\mathbf{W}^*$ will take values on the boundary of $\mc{W}$, in which case bootstrapping may fail to estimate the distribution of $\hat{\alpha}_{\mrm{wadj}}$ \citep{andrews2000inconsistency}.  When $p < n$ and there exists some $\mathbf{W}^*\in \mc{W}$ satisfies (\ref{SCM}), $\mc{B}_u$ fails since the non-uniqueness due to $p < n$ will guarantee degeneracy of  $\mathbf{W}^*$. However, this issue will not occur under $\mc{B}_f$ since it takes $\mbf{W}^*$ as being fixed and the parameter space is $\Theta$ that does not involve the constrained $\mc{W}$. Moreover, simulations in Section 4 of the Supplementary Materials show that non-uniqueness does not seriously compromise the inference.

\section{Forecasting Conoco Phillips stock in the presence of shocks}
\label{forecasting}

We demonstrate our post-shock forecasting methodology on a time series of Conoco Phillips share prices after the occurrence of a structural shock. Conoco Phillips is a large oil and gas resources company \citep{conocowhatwedo}. The particular post-shock response that we predict happened after trading ended on Friday March 6th, 2020 and before trading began on Monday March 9th, 2020. It is reasonable that the timing of this shock is known, several events occurred over the trading weekend which had an impact on stock markets and the oil markets. For example, Russia and OPEC began a battle for global oil price control on Sunday, March 8th \citep{sukhankin2020russian}, and several US states began declaring state of emergencies in response to the evolving coronavirus pandemic \citep{nygov, alonso2020state}. In this analysis we make the following design considerations:

\begin{enumerate}
\item[(1)] {\bf Selection of model}. We will use an AR(1) model to forecast Conoco Phillips stock price. This model has been shown to beat no-change forecasts when predicting oil prices over time horizons of one and three months \citep{alquist2013forecasting}.  For these reasons, we will consider 30 pre-shock trading days and we will forecast the immediate shock effect. All estimates will be adjusted for inflation. The model setup for AR(1) is exactly the same as what is stated in Section \ref{modelsetup} with addition of shock effects. All the parameters are estimated using OLS.
\item[(2)] {\bf Selection of covariates}. We consider different covariates for the model of $\alpha_i$ and $y_{i, t}$. The model of $\alpha_i$ incorporates daily S\&P 500 index prices, West Texas Intermediate (WTI) crude oil prices, dollar index, 13 Week treasury bill rates, and Chicago Board Options Exchange volatility index (VIX). The model for $y_{i,t}$ disregards VIX. This is because VIX is a metric for capturing market risk and sentiment, which is highly influential for the shock effect model.
\item[(3)] {\bf Construction of donor pool}. Our donor pool consists of Conoco Phillips shock effects observed in the past. We consider shock effects which occurred on March 14, 2008, several days in September, 2008, and November 27, 2014. The first sets of shock effects were observed during recessions that possessed similar characteristics to the current recession. In particular, all of these recessions were predicated by an inversion of the yield curve \citep{bauer2018economic}. These 2008 shock effects correspond to the collapse of Bear Stearns, the placement of Fannie May and Freddie Mac in conservatorship on September 7th, the collapse of Lehman Brothers on September 15th, and the closing of Washington Mutual on September 25th \citep{shorter2008bear, ewing2013volatility, dwyer2009financial, longstaff2010subprime}. The last shock effect corresponds to an OPEC induced supply side shock effect \citep{huppmann2015opec}. %The reasons for those three shocks are.
\end{enumerate}

We assume that the five shocks are independent of the shock that Conoco Phillips experienced on March 9, 2020. The covariates and response of time series in the donor pool are adjusted for inflation. Note that there are three shock effects nested in the time series 2008 September, we assume that these three shocks are independent, where the assumption checks using likelihood ratio test are provided in the Section 1 in the Supplementary Materials. The estimated shock-effects for $\alpha_i$ are $-0.922, -7.063, -5.777, -6.395, -4.207$ for $i = 2, \ldots, 6$, respectively. Under $\mc{M}_{2}$, we computed $\hat{\alpha}_{\rm adj}$, weighted adjustment $\hat{\alpha}_{\rm wadj}$, and $\hat{\alpha}_{\rm IVW}$. Note that non-uniqueness problems will not occur in this analysis since the conditions of Proposition \ref{uniqueness} are satisfied. To avoid the effect of  unit differences on weighting, we center and scaled the covariates in weights computation but not in the model of $y_{i,t}$. For $\hat{\alpha}_{\rm wadj}$, we observe that 
$
  \mathbf{W}^*= (0.000,0.000, 0.000, 0.273, 0.727)
$
and 
$
  \Vert\mbf{X}_1-\hat{\mbf{X}}_1(\mbf{W}^*)\Vert_2 = 3.440.
$
Note that the norm is computed using the $k$-dimensional Euclidean metric. The solution $\mathbf{W}^*$ suggests that the shock effect of interest is very similar to the September 25, 2008 shock effect and  the November 27, 2014 shock effect. 

%\begin{table}[H]
%  \caption{Bootstrap estimates and results yielded by risk-reduction propositions with $B = 1000$}\label{table5}
%  \begin{center}
%    \begin{tabular}{lrrr}
%      & $\hat{\alpha}_{\rm adj}$ & $\hat{\alpha}_{\rm wadj}$ & $\hat{\alpha}_{\rm IVW}$ \\
 %     \hline 
 %   Bootstrapped variance & 0.531 & 0.479 & 0.970 \\
 %   \end{tabular}
 % \end{center}  
%\end{table}
%\vspace{-.5cm}

The resulting shock effect estimates are $\hat{\alpha}_{\rm adj}=-4.872$, $\hat{\alpha}_{\rm wadj}= -4.805$, and $\hat{\alpha}_{\rm IVW}=-4.384$. Using the bootstrap procedure $\mc{B}_f$, we estimated parameters for risk-reduction propositions and risk-reduction quantities proposed in  Section \ref{properties}. The estimated bootstrap variances for $\hat{\alpha}_{\text{adj}}$, $\hat{\alpha}_{\text{wadj}}$, and $\hat{\alpha}_{\text{IVW}}$  are 0.419, 0.559, and 0.667, respectively. %Plugging these estimates into conditions in Section \ref{properties} yields: (1) $\hat{\alpha}_{\text{adj}}$, $\hat{\alpha}_{\text{wadj}}$, and $\hat{\alpha}_{\text{IVW}}$  reduce the risk and (2) the risk-reduction quantities are $32.452$, $32.690$, and $31.960$, respectively. 
We verify the consistency of the result yielded by risk-reduction propositions with the reality as below.

\begin{figure}
  \begin{center}
    \includegraphics[height = 9cm]{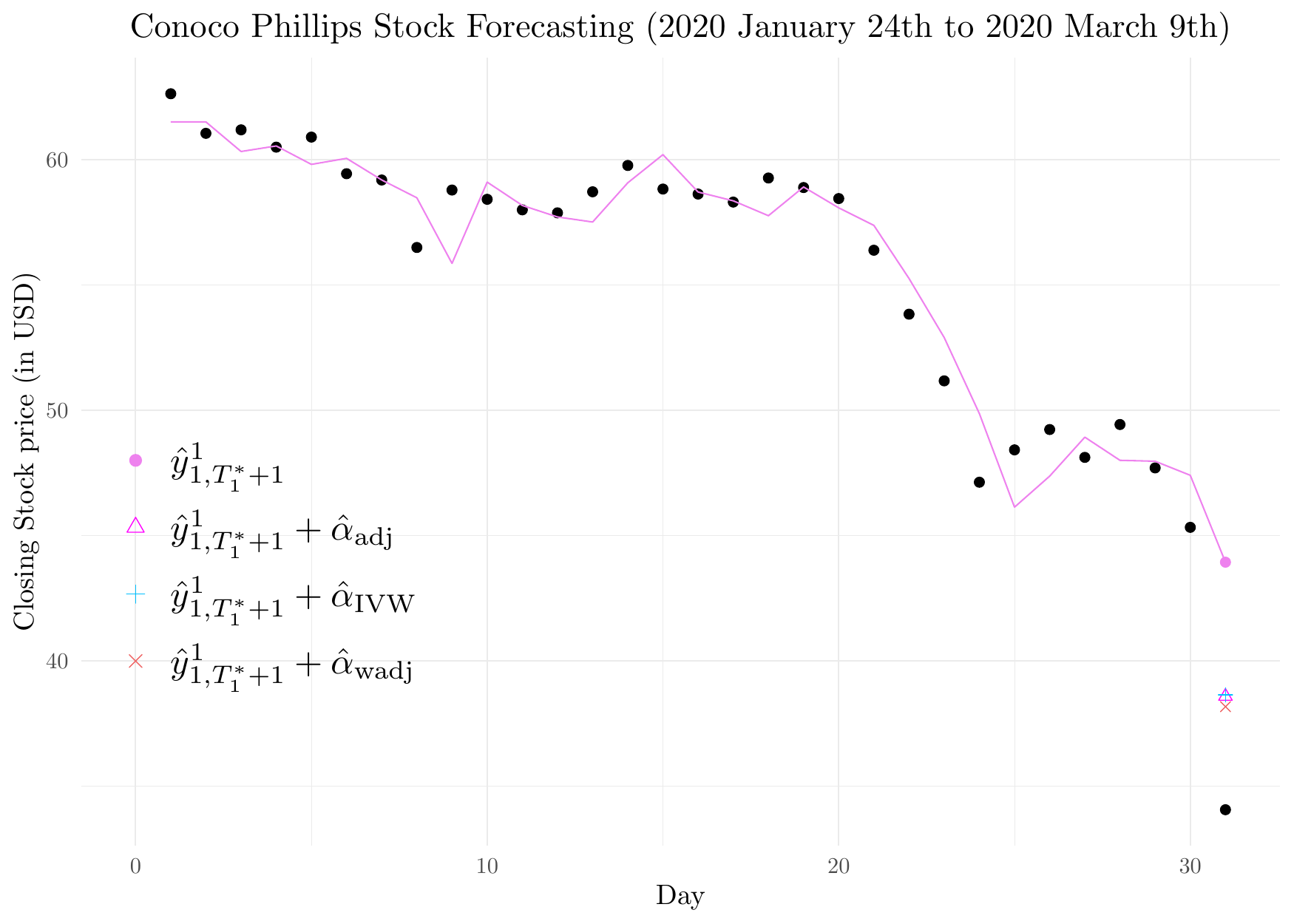}
    \caption{March 9th, 2020 post-shock forecasts for Conoco Phillips stock price.}
    \label{Fig:CP}
  \end{center}  
  \vspace{-.6cm}
\end{figure}

We can see from Figure~\ref{Fig:CP} that $\hat{\alpha}_{\text{adj}}$, $\hat{\alpha}_{\text{wadj}}$ and $\hat{\alpha}_{\text{IVW}}$ perform decently well. Our estimation framework misses the yet to be observed pot-shock response, however they perform much better than unadjusted forecasts that do not account for shock effects. The unadjusted forecast misses the post-shock response by 9.870 dollars whereas 
the use of $\hat{\alpha}_{\text{adj}}$, $\hat{\alpha}_{\text{wadj}}$, and $\hat{\alpha}_{\text{IVW}}$ misses by $5.324$, $5.392$, and $5.813$ dollars, respectively. 
%Therefore, the risk-reduction propositions are consistent with the reality  with the reduced risks for forecasts using  $\hat{\alpha}_{\text{adj}}$, $\hat{\alpha}_{\text{wadj}}$, and $\hat{\alpha}_{\text{IVW}}$ than without. Moreover, the risk-reduction quantities are consistent with the reality as well. 
The true shock effect is not fully recovered by $\hat{\alpha}_{\text{adj}}$, $\hat{\alpha}_{\text{wadj}}$, and $\hat{\alpha}_{\text{IVW}}$. This may be a result of a poorly constructed donor pool. The shock on March 9th, 2020 is in the midst of the COVID-19 pandemic and oil production volatility. It is difficult to find available stock market time series data that were generated under a similar setting. In any event, the shock on March 9th, 2020 was the largest price shock to Conoco Phillips shares by a wide margin, even after adjusting for inflation.

From another perspective, it is possible that the stock of Conoco Phillips actually experienced multiple shocks on 2020 March 9th. For example, \citet{kilian2009not} studied the effect that different supply and demand shocks have on oil prices through a vector autoregressive model. Their model postulates an additive nature of shock effects, although the additivity parameters requires estimation in their context. Motivated by \citet{kilian2009not}, we also studied additive shock effect estimators where the shock effects corresponding to separate supply and demand shocks are added to estimate the unknown shock effect. The supply shock donor pool consists of the November 27, 2014 shock effect; and the demand shock donor pool consists of the remaining shock effects. The additive adjustment estimator computed by adding the $\hat{\alpha}_{\text{adj}}$, $\hat{\alpha}_{\text{wadj}}$, and $\hat{\alpha}_{\text{IVW}}$ estimators for the demand and supply shock effects only, respectively, miss the post-shock Conoco Philips share price by $0.951$, $0.405$, and $1.460$ dollars. These additive adjustment estimators do extremely well in this additive shock setting. 
%There is apriori justification for the use of these simply additive adjustment estimators, although their nearly perfect performance in this example is a retrospective finding.  %This provides some anecdotal justification for apriori estimation of the March 9th Conoco Phillips shock effect with the addition of the adjustment estimator corresponding to the separate supply and demand aspects of this shock effect.

\section{Discussion}
\label{discussion}

We developed a methodology for forecasting post-shock response values after the occurrence of a structural shock. Our methodology is as follows: construct a synthetic panel of time series which have undergone similar shocks, estimate the shock effects in those series, aggregate them, and then adjust the original forecast by adding the aggregated shock effect estimator to the original forecast. There have been several other recent similar methods developed for forecasting COVID-19 cases. For example, \citet{lee2020estimation} constructed a Bayesian hierarchical model embracing data integration to improve predictive precision of COVID-19 infection trajectories for different countries. A similar setup may be appropriate for post-shock forecasting but may be too dependent upon model specification for the shock distribution. \citet{plessen2020integrated} employed a data-mining approach to combine COVID-19 data from different countries as input to predict global net daily infections and deaths of COVID-19 using a clustering approach. However, there is a tremendous amount of volatility in this form of COVID-19 data, and the fit of this prediction method may be improved with modeling structure or preprocessing of the donor pool. \citet{agarwal2020two} proposed a model-free synthetic intervention method to predict unobserved potential outcomes after different interventions given a donor pool of observed outcomes with given interventions. They also provided useful guidelines for how to estimate the effects of potential interventions by giving recommendations for choosing the metric of interest, the intervention of interest, time horizons, and the donor pool. Although the methodology in \citet{agarwal2020two} is quite general, there is no guarantee for theoretical properties in prediction without assuming any distributional structure.

We provided risk-reduction propositions and empirical tools that can prospectively assess the effectiveness of our adjustment strategies in additive shock effect settings. The model, under which we verify these claims, is a simple AR(1) model. Similar results can be obtained for more general models such as AR($p$), vector autoregression, and generalized autoregressive conditional heteroskedasticity models. Generally, multiple shock effects can be nested within a time series; and time series in the donor pool can be dependent. As an example, we  considered a dependency structure for the September 2008 shock effects in our analysis of Conoco Phillips stock. But we note that consistency estimates from LOOCV may not work well if donor pool candidates are not mutually independent since the almost unbiased property hinges on the mutual independence among candidates in the donor pool.
%This generalization is possible since our risk-reduction propositions in Section \ref{properties} assumes the shock effect estimates are independent of $\alpha_1$, the shock effect for the time series of interest; and allows arbitrary dependence of shock effects in the donor pool. 
%Specific to this generalization, we edit the model of $\alpha_i$ to $\alpha_i = \mu_{\alpha} + \delta_i'\mathbf{x}_{3, T_i^* + 1} + \tilde{\varepsilon}_i$ for $i = 3, 4, 5$ and $\alpha_i = \mu_{\alpha} + \delta_i'\mathbf{x}_{4, T_i^* + 1} + \tilde{\varepsilon}_i$ for $i = 6$.
%This generalization was adopted in Section \ref{forecasting}. In this case, the variance expressions proposed in Section \ref{properties} will not work. However, the risk-reduction propositions in Section \ref{properties} will work  as long as those shock effects (not necessarily mutually independent) in the donor pool are independent of the one of interest. 
Although it is reflected in $\mc{M}_2$, we stress that our proposed methods allow $\alpha_i$ to follow arbitrary distributions provided that its first and second moments exist. The covariates in the model for $\alpha_i$ under $\mc{M}_2$ can be different from the covariates in the model of $y_{i,t}$.  Additionally, $\alpha_i$ can be heteroskedastic; i.e.,  they can have different variances. In this scenario, all the theoretical properties still hold though variance expressions in Section \ref{conditions} may not apply. We also note that our post-shock framework can be extended to settings where the shock effect can be decomposed into separable estimable parts. An example of this is the additive shock effect estimators that we studied in our Conoco Phillips analysis. Although our work is developed for time-series or AR$(p)$ models, in fact, it can be generalized to any similar setting with a model of the response, whose parameters can be estimated unbiasedly, an additive shock-effect structure, and the structure that the time series in the donor pool are independent of the one of interest.

Our bootstrap procedures can be extended to approximate the distribution of shock effect estimators from more general time series. If the data are subject to heteroskedasticity of unknown form, bootstrapping tuples of regressands and regressors proposed by \citet{freedman1981bootstrapping} is robust in this situation with asymptotic validity in autoregressive models established by \citet{gonccalves2004bootstrapping}. If serial correlation exists in the data, various block bootstrapping procedures \citep{kunsch1989jackknife, liu1992moving} can be possible reasonable alternatives. Note that the pseudo time series generated by our proposed residual bootstrap are not stationary. If stationarity is of concern, one can be referred to the stationary bootstrap in \citet{politis1994stationary} for stationary and weakly dependent time series. Nevertheless, it was shown that approximation accuracy might be a cost for the stationary bootstrap in autoregressive  models in finite sample \citep{berkowitz1999finite}.  More work related to bootstrapping time series  can be referred to Chapters 3 and 4 in \citet{politis1999subsampling}, \citet{berkowitz2000recent}, and Chapter 12 in \citet{kilian2017structural}. It is up to  users in terms of selecting which procedure to choose but under different assumptions on the time series.

% \citet{politis1994stationary} motivated a stationary bootstrap method for strictly stationary and weakly dependent time series $\{X_n \colon n \in \mathbb{N}\}$.  This algorithm generates a sequence of blocks of observations $B_{I_1, L_1}, B_{I_2, L_2}, \ldots$ where $B_{i,b}= \{X_i, X_{i+1}, \ldots, X_{i+b-1}\}$; for $j > N$, $X_{j}$ is defined to be $X_k$, where $k = j \hspace{-.1cm} \mod N$ and $X_0 = X_N$. The sampling stops when $N$ observations are reached. Note that the collection of random positions $\{I_n \colon n \in \mathbb{N}\}$ is a sequence of i.i.d. discrete uniform random variables; and the collection of random lengths $\{L_n \colon n \in \mathbb{N}\}$ is a sequence of  i.i.d. geometric random variables with parameter $p$.  However, the consistency needs to be proved by a case-by-case analysis \citep[Page 66]{politis1999subsampling}. Additionally, the asymptotic accuracy of this algorithm can be sensitive to the selection of $p$. This issue is similar to that of the selection of block size in moving-block bootstrapping \citep{kunsch1989jackknife, liu1992moving}. 

Construction of donor pool is a critical step that directly matters in prediction. In our model, shocks from candidate time series in the donor pool, together with the shock in the time series of interest, should come from a family of distributions that allow for  varying variances and  varying means that are linear functions of common covariates. Remark \ref{remark1} shows that common covariates are not restrictive with structural zero in coefficients. However, it is difficult to verify such assumption or structure in practice. In principle, one should select identities that come from a common population, which is also the one  the identity under study come from, as candidates in donor pool. This selection criterion is often adopted in the literature of comparative case studies. For example, \citet{card1990impact} used a set of cities in southern U.S. to estimate the effect of 1980 Mariel Boatlift on the Miami labor market. \citet{abadie2003economic} utilized a donor pool of two Spanish regions to approximate the effect of terrorism on  the economic growth of Baseque Country. To estimate the effect of Proposition 99, a large tabacco control program in California  in 1988, on the annual per-capita cigarette sales of 2000, \cite{abadie2010synthetic} constructed a donor pool consisting of US states, which did not implement \emph{any} large-scale tobacco control program during their sampling period. Common features of those studies are that they all carefully constructed the donor pool and discarded any suspicious outliers.

\vspace*{0.5cm}\noindent{\bf Acknowledgements}: We are grateful to Forrest W. Crawford, Karl Oskar Ekvall, Soheil Eshghi, Lutz Kilian, Ziyu Liu, Dootika Vats, and Dave Zhao for helpful comments.

\section{Appendix}
\label{proofs}

\subsection{Justification of Expectation of $\hat{\alpha}_{\rm adj}$ and $\hat{\alpha}_{\rm wadj}$}
\label{exp}

The building block for the following proof is the fact that least squares is conditionally unbiased conditioned on $\Theta$. 

\noindent \textbf{Case I: under $\mc{M}_{1}$:} It follows that  under $\mc{M}_{1}$ (see Section \ref{modelsetup}),
\begin{align*}
\E{\hat{\alpha}_{\rm adj}} =\frac{1}{n}  \sum_{i=2}^{n+1} \E{\E{\hat{\alpha}_i|\Theta}} = \mu_{\alpha} 
\quad \text{ and } \quad \E{\hat{\alpha}_{\rm wadj}}&= \sum_{i=2}^{n+1} w_i^* \E{\E{\hat{\alpha}_i|\Theta}}= \sum_{i=2}^{n+1} w_i^*\mu_{\alpha}= \mu_{\alpha}.
\end{align*}
where we used the fact that $\sum_{i=2}^{n+1} w^*_i=1$. 

\noindent \textbf{Case II: under $\mc{M}_{21}$ and $\mc{M}_{22}$:} Since $\E{\tilde{\varepsilon}_{i, T_i}}=0$, $\E{\hat{\tilde{\alpha}}_{i}}=\E{\tilde{\alpha}_{i}}=\E{\alpha_{i}}$, it follows that
  \begin{align*}
   \E{\hat{\alpha}_{\rm wadj}}= \mrm{E}\left\{\mrm{E}\left(\sum_{i=2}^{n+1} w_i^*\hat{\alpha}_i|\Theta\right)\right\}
   &= \mrm{E}\left( \sum_{i=2}^{n+1} w_i^*\alpha_i\right)\\
   &= \mrm{E}\bigg\{ \sum_{i=2}^{n+1} w_i^*\left[\mu_{\alpha}+\delta_{i}'\mbf{x}_{i, T_i^*+1}\right]\bigg\}\\
   &=\mu_{\alpha}+ \mu_{\delta}' \sum_{i=2}^{n+1} w_i^*\mbf{x}_{i, T_i^*+1}\tag{$\mbf{W}\in \mc{W}$}
   \\
   &=\mu_{\alpha}+  \mu_{\delta}'\mbf{x}_{1, T_1^*+1}. \tag{from (\ref{SCM})}
   \end{align*}
Similarly,
  \begin{align*}
   \E{\hat{\alpha}_{\rm adj}}
   &=\mu_{\alpha}+\frac{1}{n}\sum_{i=2}^{n+1} \mu_{\delta}'  \mbf{x}_{i, T_i^*+1}.
   \end{align*}

\subsection{Justification of Variance of $\hat{\alpha}_{\rm adj}$ and $\hat{\alpha}_{\rm wadj}$}
\label{var}

Notice that under the setting of OLS, the design matrix for $\mc{M}_2$ is the same as the one for $\mc{M}_1$. Therefore, it follows that
  \begin{align*}
  \var{\hat{\alpha}_{\rm wadj}} 
  &= \E{\var{\hat{\alpha}_{\rm wadj}|\Theta}} + \var{\E{\hat{\alpha}_{\rm wadj}|\Theta}} \\
  &=\mrm{E} \left\{\mrm{Var}\left(\sum_{i=2}^{n+1} w_i^*\hat{\alpha}_i|\Theta\right)\right\} +\mrm{Var}\left(\sum_{i=2}^{n+1} w_i^*\alpha_i\right) 
\end{align*}
Under $\mc{M}_{21}$ where $\delta_i=\delta$ are fixed unknown parameters,  we will have
  \begin{align}
  \var{\hat{\alpha}_{\rm wadj}} 
  &= \mrm{E} \left\{\sum_{i=2}^{n+1}(w_i^*)^2(\sigma^2(\mbf{U}'_i\mbf{U}_i)^{-1}_{22})\right\} +\sigma^2_{\alpha}\sum_{i=2}^{n+1}(w_i^*)^2  \nonumber\\
  &= \sigma^2\sum_{i=2}^{n+1}(w_i^*)^2\mrm{E}\big\{(\mbf{U}'_i\mbf{U}_i)^{-1}_{22}\big\}+\sigma^2_{\alpha}\sum_{i=2}^{n+1}(w_i^*)^2.\label{equation6}
\end{align}
Similarly, under $\mc{M}_{22}$ where we assume $\delta_i \indep \varepsilon_{i,t}$, we have
 \begin{align*}
  \var{\hat{\alpha}_{\rm wadj}} 
  &= \sigma^2\sum_{i=2}^{n+1}(w_i^*)^2\mrm{E}\big\{(\mbf{U}'_i\mbf{U}_i)^{-1}_{22}\big\}
  + \sum_{i=2}^{n+1} (w_i^*)^2 (\mbf{x}_{i, T_i^*+1}'\Sigma_{\delta}\mbf{x}_{i, T_i^*+1} + \sigma^2_{\alpha})
\end{align*}
For the adjustment estimator, we simply replace $\mbf{W}^*$ with $1/n\mbf{1}_n$. Thus, under $\mc{M}_{21}$ we have 
 \begin{align*}
  \var{\hat{\alpha}_{\rm adj}} 
  &=\frac{\sigma^2}{n^2}\sum_{i=2}^{n+1}\mrm{E}\big\{(\mbf{U}'_i\mbf{U}_i)^{-1}_{22}\big\}+\frac{\sigma^2_{\alpha}}{n^2}
\end{align*}
Under $\mc{M}_{22}$, we shall have
 \begin{align*}
  \var{\hat{\alpha}_{\rm adj}} 
  &=\frac{\sigma^2}{n^2}\sum_{i=2}^{n+1}\mrm{E}\big\{(\mbf{U}'_i\mbf{U}_i)^{-1}_{22}\big\}+\frac{1}{n^2}(\mbf{x}_{i, T_i^*+1}'\Sigma_{\delta}\mbf{x}_{i, T_i^*+1} + \sigma^2_{\alpha}).
\end{align*}
Notice that $\mc{M}_{1}$ differs from $\mc{M}_{21}$ only by its mean parameterization of $\alpha$ (see Section \ref{modelsetup}). In other words, the variances of $\hat{\alpha}_{\rm adj}$ and $\hat{\alpha}_{\rm wadj}$ under $\mc{M}_1$ are the same for those under $\mc{M}_{21}$.

\subsection{Proofs for lemmas and propositions}

\begin{proof-of-proposition}[\ref{uniqueness}]
  The proof of \citet{li2019statistical} in Appendix A.2 and A.3 adapts easily to Proposition \ref{uniqueness}.
\end{proof-of-proposition}

\begin{proof-of-proposition}[\ref{unbiased}] The proof for unbiasedness follows immediately from discussions related to expectation in Section \ref{properties}. For the biasedness of  $\hat{\alpha}_{\rm adj}$ under $\mc{M}_{21}$ and $\mc{M}_{22}$, we write the bias term for $\hat{\alpha}_{\rm adj}$ as below.
\begin{align*}
  \mrm{Bias}(\hat{\alpha}_{\rm adj}) = 
  \begin{cases}
       \frac{1}{n} \sum_{i=2}^{n+1} \delta'(\mbf{x}_{i, T_i^*+1}-n\mbf{x}_{1, T_1^*+1})  & \text{ for } \mc{M}_{21}\\
    \frac{1}{n} \sum_{i=2}^{n+1} \mu_{\delta}'(\mbf{x}_{i, T_i^*+1}-n\mbf{x}_{1, T_1^*+1})  & \text{ for }\mc{M}_{22}
  \end{cases}.
\end{align*}
But it may be unbiased in some special circumstances when the above bias turns out to be 0. \end{proof-of-proposition}

\begin{lem}
  \label{risklemma} The forecast risk reduction is $R_{T_1^*+1,1}-R_{T_1^*+1,2}=\E{\alpha_1^2}-\E{\hat{\alpha}-\alpha_1}^2$ for all estimators of $\alpha_1$ that are independent of $\Theta_1$ (see Section \ref{modelsetup}).
\end{lem}

\begin{proof-of-lemma}[\ref{risklemma}]
  Define 
  \begin{align*}
    C(\Theta_1) =\hat{\eta}_1 +\hat{\phi}_1 y_{1, T_1^*}+\hat{\theta}_1'\mbf{x}_{1, T_1^*+1} -(\eta_1 +\phi_1y_{1,T_1^*}+\theta_1'\mbf{x}_{1,T_1^*+1}),
  \end{align*}
  where $\Theta_1$ is as defined in (\ref{parameter}). Notice that
  \begin{align*}
    R_{T_1^*+1,1}= \mrm{E}\big\{\big(C(\Theta_1)-\alpha_1\big)^2\big\}
    \qquad \text{ and } 
    \qquad  R_{T_1^*+1,2}= \mrm{E}\big\{\big(C(\Theta_1)+\hat{\alpha}-\alpha_1\big)^2\big\}.
  \end{align*}
  It follows that
  \begin{align*}
    R_{T_1^*+1,1}-R_{T_1^*+1,2}=\E{\alpha_1^2}-2\E{C(\Theta_1)\hat{\alpha}}-\E{\hat{\alpha}-\alpha_1}^2.
  \end{align*}
  Assuming $\mbf{S}=(\mbf{1}_n, \mbf{y}_{1,t-1}, \mbf{x}_{1})$ has full rank, under OLS setting, $\hat{\eta}_1$, $\hat{\phi}_1$, and $\hat{\theta}_1$ are unbiased estimators of $\eta_1$, $\phi_1$, and $\theta_1$, respectively under conditioning of $\Theta_1$. Since we assume $\hat{\alpha}$ is independent of $\Theta_1$, through the method of iterated expectation,
  \begin{align*}
    \E{C(\Theta_1)\hat{\alpha}}=\mrm{E}\big\{\hat{\alpha}\cdot \E{C(\Theta_1)\mid \Theta_1}\}=0.
  \end{align*}
  It follows that
  \begin{align*}
    R_{T_1^*+1,1}-R_{T_1^*+1,2}=\E{\alpha_1^2}-\E{\hat{\alpha}-\alpha_1}^2,
  \end{align*}
  which finishes the proof.
\end{proof-of-lemma}

\begin{proof-of-proposition}[\ref{proprisk}] The proofs are arranged into two separate parts as below.

 \textbf{Proof for statement (i):} Under $\mc{M}_1$, $\hat{\alpha}_{\rm adj}$ is an unbiased estimator of $\E{\alpha_1}$ because
  \begin{align*}
   \mrm{E}\left( \frac{1}{n}\sum_{i=2}^{n+1} \hat{\alpha}_i\right)
   = \frac{1}{n}\sum_{i=2}^{n+1}\E{\hat{\alpha}_i}
   &= \frac{1}{n}\sum_{i=2}^{n+1}\E{\E{\hat{\alpha}_i\mid \Theta}}\\
   &=  \frac{1}{n}\sum_{i=2}^{n+1}\E{\alpha_i}
   = \mu_{\alpha}=\E{\alpha_1},
  \end{align*}
  where we used the fact that OLS estimator is unbiased when the design matrix $\mbf{U}_i$ is of full rank for all $i = 2, \ldots, n+1$. Because $\alpha_1\indep \varepsilon_{i,t}$, $\E{\hat{\alpha}_{\rm adj}\alpha_1}=\E{\hat{\alpha}_{\rm adj}}\E{\alpha_1}=(\E{\hat{\alpha}_{\rm adj}})^2$. By Lemma \ref{risklemma}, 
    \begin{align*}
    R_{T_1^*+1,1}-R_{T_1^*+1,2}
    &=\E{\alpha_1^2}-\E{\hat{\alpha}_{\rm adj}-\alpha_1}^2\\
   & =\E{\alpha_1^2}-\E{\alpha_1^2}- \E{\hat{\alpha}_{\rm adj}^2}+2\E{\hat{\alpha}_{\rm adj}\alpha_1} \\
   &= \mu_{\alpha}^2 - \var{\hat{\alpha}_{\rm adj}} 
  \end{align*}
  Therefore, as long as we have $\var{\hat{\alpha}_{\rm adj}}<\mu_{\alpha}^2$, we will achieve the risk reduction. 

 \textbf{Proof for statement (ii):} By Proposition \ref{unbiased}, the property that $\hat{\alpha}_{\rm wadj}$ is an unbiased estimator of $\mu_{\alpha}$ holds for $\mc{M}_{1}$. The remainder of the proof follows a similar argument to the proof of statement (i).
\end{proof-of-proposition}

\begin{proof-of-proposition}[\ref{propriskwadj2}]
  By Proposition \ref{unbiased}, the property that $\hat{\alpha}_{\rm wadj}$ is an unbiased estimator of $\E{\alpha_1}$ holds for $\mc{M}_{21}$ and $\mc{M}_{22}$. The remainder of the proof follows a similar argument to the proof of Proposition \ref{proprisk}.
\end{proof-of-proposition}

\subsection{Tables for Section \ref{simulation}}
\label{tablesappendix}
\newpage

\begin{landscape}
\begin{table}[b]
\caption{30 Monte Carlo simulations of $\mc{M}_2$ for $\mc{B}_u$ with varying $n$ and $\sigma_{\alpha}$} \vspace{.3cm} \label{table1}
\begin{center}
\resizebox{1.4\textwidth}{!}{\begin{tabular}{cc|ccc|ccc|cccc|}
   &   & \multicolumn{3}{|c|}{Guess} & \multicolumn{3}{|c|}{LOOCV with $k$ random draws} &  \multicolumn{4}{|c|}{Distance to $y_{1, T_1^*+1}$} \\ 
 $n$   & $\sigma_{\alpha}$ &  $\delta_{\hat{\alpha}_{\rm adj}}$  & $\delta_{\hat{\alpha}_{\rm wadj}}$ & $\delta_{\hat{\alpha}_{\rm IVW}}$  & $\bar{\mc{C}}^{(k)}(\delta_{\hat{\alpha}_{\rm adj}})$  & $\bar{\mc{C}}^{(k)}(\delta_{\hat{\alpha}_{\rm wadj}})$ & $\bar{\mc{C}}^{(k)}(\delta_{\hat{\alpha}_{\rm IVW}})$ & Original & $\hat{\alpha}_{\rm adj}$ & $\hat{\alpha}_{\rm wadj}$ & $\hat{\alpha}_{\rm IVW}$\\[.15cm] 
  \hline
 \multirow{5}{*}{5} & 5  & 1 (0) & 1 (0) & 1 (0) & 0.91 (0.03) & 0.91 (0.02) & 0.9 (0.03) & 53.23 (4.1) & 15.88 (2.1) & 16.78 (2.37) & 15.82 (2.07) \\ 
    & 10  & 0.97 (0.03) & 1 (0) & 0.97 (0.03) & 0.89 (0.03) & 0.9 (0.03) & 0.89 (0.03) & 53.01 (4.47) & 17.83 (2.38) & 19.56 (2.56) & 17.61 (2.36) \\ 
    & 25  & 0.93 (0.05) & 0.97 (0.03) & 0.93 (0.05) & 0.74 (0.04) & 0.81 (0.04) & 0.75 (0.04) & 53.38 (5.92) & 26.44 (3.8) & 29.06 (4) & 26.11 (3.75) \\ 
    & 50  & 0.83 (0.07) & 0.83 (0.07) & 0.8 (0.07) & 0.59 (0.05) & 0.64 (0.05) & 0.59 (0.05) & 61.68 (7.73) & 46 (6.31) & 47.3 (7.14) & 45.25 (6.32) \\ 
    & 100  & 0.7 (0.09) & 0.87 (0.06) & 0.7 (0.09) & 0.53 (0.05) & 0.54 (0.05) & 0.53 (0.06) & 85.68 (12.95) & 87.25 (11.86) & 87.07 (13.63) & 85.65 (12.02)\\[.3cm] 
   \multirow{5}{*}{10} & 5  & 1 (0) & 1 (0) & 1 (0) & 0.91 (0.03) & 0.92 (0.02) & 0.91 (0.03) & 48.18 (4.59) & 20.47 (2.71) & 19.13 (2.97) & 20.53 (2.73) \\ 
    & 10  & 1 (0) & 1 (0) & 1 (0) & 0.87 (0.03) & 0.89 (0.03) & 0.87 (0.03) & 48.93 (4.71) & 21.27 (2.6) & 19.24 (3.03) & 21.31 (2.61) \\ 
    & 25  & 0.93 (0.05) & 0.97 (0.03) & 0.9 (0.06) & 0.74 (0.03) & 0.77 (0.03) & 0.74 (0.03) & 51.18 (5.7) & 26.68 (2.78) & 27.17 (3) & 26.53 (2.77) \\ 
    & 50  & 0.8 (0.07) & 0.8 (0.07) & 0.8 (0.07) & 0.57 (0.04) & 0.61 (0.04) & 0.57 (0.04) & 57.82 (7.81) & 40.51 (4.37) & 46.85 (4.02) & 40.19 (4.27) \\ 
    & 100  & 0.73 (0.08) & 0.93 (0.05) & 0.7 (0.09) & 0.51 (0.04) & 0.51 (0.04) & 0.5 (0.04) & 79.3 (12.44) & 72.33 (9.12) & 88.81 (8.46) & 71.83 (8.85) \\[.3cm] 
   \multirow{5}{*}{15} & 5  & 1 (0) & 1 (0) & 1 (0) & 0.94 (0.02) & 0.95 (0.02) & 0.94 (0.02) & 51.11 (3.05) & 14.94 (2.36) & 14.09 (2.37) & 15.09 (2.35) \\ 
    & 10  & 1 (0) & 1 (0) & 1 (0) & 0.92 (0.02) & 0.91 (0.02) & 0.91 (0.02) & 52.34 (3.19) & 15.64 (2.73) & 15.29 (2.74) & 15.89 (2.68) \\ 
    & 25  & 0.93 (0.05) & 0.97 (0.03) & 0.93 (0.05) & 0.73 (0.04) & 0.76 (0.04) & 0.73 (0.04) & 56.03 (5.2) & 25.49 (3.85) & 27.38 (3.87) & 25.27 (3.84) \\ 
    & 50  & 0.8 (0.07) & 0.83 (0.07) & 0.8 (0.07) & 0.56 (0.04) & 0.6 (0.04) & 0.57 (0.04) & 71.37 (7.76) & 47.25 (6.42) & 52.06 (6.57) & 46.41 (6.43) \\ 
    & 100  & 0.63 (0.09) & 0.67 (0.09) & 0.63 (0.09) & 0.52 (0.04) & 0.42 (0.04) & 0.53 (0.04) & 111.91 (13.83) & 92.95 (12.34) & 103.13 (12.74) & 91.07 (12.37) \\[.3cm] 
   \multirow{5}{*}{25} & 5  & 1 (0) & 1 (0) & 1 (0) & 0.93 (0.02) & 0.94 (0.02) & 0.93 (0.02) & 47.79 (2.93) & 14.83 (1.72) & 14.83 (2.04) & 14.76 (1.72) \\ 
    & 10  & 1 (0) & 1 (0) & 1 (0) & 0.89 (0.03) & 0.91 (0.02) & 0.89 (0.03) & 47.93 (3.25) & 16.55 (1.89) & 17.55 (2.12) & 16.53 (1.88) \\ 
    & 25  & 1 (0) & 1 (0) & 1 (0) & 0.83 (0.03) & 0.82 (0.03) & 0.83 (0.03) & 49.78 (5.01) & 26.42 (3.38) & 29.11 (3.4) & 26.45 (3.35) \\ 
    & 50  & 0.97 (0.03) & 1 (0) & 0.93 (0.05) & 0.64 (0.05) & 0.63 (0.05) & 0.64 (0.05) & 62.64 (7.4) & 48.84 (6.4) & 52.67 (6.28) & 48.8 (6.35) \\ 
    & 100  & 0.83 (0.07) & 0.8 (0.07) & 0.83 (0.07) & 0.57 (0.05) & 0.59 (0.05) & 0.59 (0.05) & 103.37 (12.23) & 97.81 (12.52) & 102.4 (12.49) & 97.63 (12.45) \\
\end{tabular}}
   \end{center}
\end{table}
\end{landscape}

\begin{landscape}
\begin{table}[b]
\caption{30 Monte Carlo simulations of $\mc{M}_2$ for $\mc{B}_u$ with varying $\sigma$ and $\sigma_{\alpha}$} \vspace{.3cm} \label{table2}
\begin{center}
\resizebox{1.4\textwidth}{!}{\begin{tabular}{cc|ccc|ccc|cccc|}
   &   & \multicolumn{3}{|c|}{Guess} & \multicolumn{3}{|c|}{LOOCV with $k$ random draws} &  \multicolumn{4}{|c|}{Distance to $y_{1, T_1^*+1}$} \\ 
 $\sigma$   & $\sigma_{\alpha}$ &  $\delta_{\hat{\alpha}_{\rm adj}}$  & $\delta_{\hat{\alpha}_{\rm wadj}}$ & $\delta_{\hat{\alpha}_{\rm IVW}}$  & $\bar{\mc{C}}^{(k)}(\delta_{\hat{\alpha}_{\rm adj}})$  & $\bar{\mc{C}}^{(k)}(\delta_{\hat{\alpha}_{\rm wadj}})$ & $\bar{\mc{C}}^{(k)}(\delta_{\hat{\alpha}_{\rm IVW}})$ & Original & $\hat{\alpha}_{\rm adj}$ & $\hat{\alpha}_{\rm wadj}$ & $\hat{\alpha}_{\rm IVW}$\\[.15cm] 
  \hline
 \multirow{5}{*}{5} & 5  & 1 (0) & 1 (0) & 1 (0) & 0.94 (0.02) & 0.97 (0.01) & 0.95 (0.02) & 48.84 (3.4) & 15.72 (1.93) & 15.32 (1.94) & 15.72 (1.93) \\ 
    & 10  & 1 (0) & 1 (0) & 1 (0) & 0.92 (0.03) & 0.94 (0.02) & 0.92 (0.03) & 49.54 (3.73) & 17.07 (2.01) & 16.08 (2.16) & 17.12 (1.99) \\ 
    & 25  & 0.87 (0.06) & 1 (0) & 0.87 (0.06) & 0.77 (0.02) & 0.81 (0.03) & 0.77 (0.03) & 51.78 (5.22) & 24.78 (2.57) & 26.12 (2.49) & 24.62 (2.54) \\ 
    & 50  & 0.8 (0.07) & 0.83 (0.07) & 0.8 (0.07) & 0.59 (0.05) & 0.61 (0.04) & 0.59 (0.05) & 58.62 (7.74) & 40.09 (4.7) & 47.09 (4.09) & 39.85 (4.58) \\ 
    & 100  & 0.7 (0.09) & 0.93 (0.05) & 0.73 (0.08) & 0.5 (0.04) & 0.49 (0.04) & 0.51 (0.04) & 82.66 (12.19) & 72.83 (9.75) & 89.03 (9.15) & 72.31 (9.5) \\[.3cm] 
   \multirow{5}{*}{10} & 5  & 1 (0) & 1 (0) & 1 (0) & 0.91 (0.03) & 0.92 (0.02) & 0.91 (0.03) & 48.18 (4.59) & 20.47 (2.71) & 19.13 (2.97) & 20.53 (2.73) \\ 
    & 10  & 1 (0) & 1 (0) & 1 (0) & 0.87 (0.03) & 0.89 (0.03) & 0.87 (0.03) & 48.93 (4.71) & 21.27 (2.6) & 19.24 (3.03) & 21.31 (2.61) \\ 
    & 25  & 0.93 (0.05) & 0.97 (0.03) & 0.9 (0.06) & 0.74 (0.03) & 0.77 (0.03) & 0.74 (0.03) & 51.18 (5.7) & 26.68 (2.78) & 27.17 (3) & 26.53 (2.77) \\ 
    & 50  & 0.8 (0.07) & 0.8 (0.07) & 0.8 (0.07) & 0.57 (0.04) & 0.61 (0.04) & 0.57 (0.04) & 57.82 (7.81) & 40.51 (4.37) & 46.85 (4.02) & 40.19 (4.27) \\ 
    & 100  & 0.73 (0.08) & 0.93 (0.05) & 0.7 (0.09) & 0.51 (0.04) & 0.51 (0.04) & 0.5 (0.04) & 79.3 (12.44) & 72.33 (9.12) & 88.81 (8.46) & 71.83 (8.85) \\[.3cm] 
   \multirow{5}{*}{25} & 5  & 0.97 (0.03) & 1 (0) & 0.97 (0.03) & 0.7 (0.04) & 0.76 (0.03) & 0.7 (0.04) & 50.09 (8.27) & 38.1 (5.75) & 37.98 (5.5) & 38.44 (5.76) \\ 
    & 10  & 0.97 (0.03) & 1 (0) & 0.97 (0.03) & 0.69 (0.04) & 0.74 (0.03) & 0.69 (0.04) & 50.82 (8.15) & 37.63 (5.61) & 36.33 (5.64) & 37.99 (5.61) \\ 
    & 25  & 0.9 (0.06) & 0.9 (0.06) & 0.9 (0.06) & 0.62 (0.04) & 0.64 (0.04) & 0.61 (0.04) & 53.01 (8.22) & 38.78 (5.29) & 36.82 (5.76) & 38.88 (5.31) \\ 
    & 50  & 0.8 (0.07) & 0.8 (0.07) & 0.8 (0.07) & 0.53 (0.04) & 0.53 (0.04) & 0.53 (0.04) & 58.9 (9.12) & 46.77 (5.54) & 50.61 (5.71) & 46.6 (5.51) \\ 
    & 100  & 0.7 (0.09) & 0.9 (0.06) & 0.67 (0.09) & 0.51 (0.03) & 0.56 (0.04) & 0.5 (0.03) & 79.64 (12.21) & 72.76 (8.79) & 89.48 (7.98) & 72.17 (8.59) \\[.3cm] 
   \multirow{5}{*}{50} & 5  & 0.77 (0.08) & 0.8 (0.07) & 0.77 (0.08) & 0.6 (0.05) & 0.63 (0.04) & 0.59 (0.04) & 71.22 (13) & 70.31 (10.4) & 72.3 (9.26) & 70.79 (10.45) \\ 
    & 10  & 0.77 (0.08) & 0.77 (0.08) & 0.77 (0.08) & 0.6 (0.05) & 0.63 (0.05) & 0.6 (0.05) & 70.85 (12.91) & 69.43 (10.22) & 70.65 (9.29) & 69.94 (10.26) \\ 
    & 25  & 0.7 (0.09) & 0.73 (0.08) & 0.7 (0.09) & 0.54 (0.05) & 0.56 (0.05) & 0.55 (0.05) & 70.32 (12.81) & 67.61 (9.86) & 67 (9.58) & 68.06 (9.89) \\ 
    & 50  & 0.67 (0.09) & 0.7 (0.09) & 0.67 (0.09) & 0.51 (0.05) & 0.51 (0.04) & 0.51 (0.05) & 74.01 (12.66) & 68.69 (9.63) & 67.9 (10.21) & 68.91 (9.64) \\ 
    & 100  & 0.5 (0.09) & 0.6 (0.09) & 0.47 (0.09) & 0.47 (0.05) & 0.49 (0.04) & 0.45 (0.05) & 92.71 (13.06) & 83.66 (10.79) & 94.53 (11.2) & 83.56 (10.63)\\[.3cm] 
   \multirow{5}{*}{100} & 5  & 0.47 (0.09) & 0.47 (0.09) & 0.47 (0.09) & 0.51 (0.06) & 0.57 (0.05) & 0.49 (0.06) & 130.47 (22.59) & 135.16 (19.73) & 141.42 (16.98) & 136.3 (19.72) \\ 
    & 10  & 0.47 (0.09) & 0.47 (0.09) & 0.47 (0.09) & 0.51 (0.05) & 0.53 (0.05) & 0.51 (0.06) & 129.49 (22.49) & 134.09 (19.52) & 139.69 (16.96) & 135.26 (19.51) \\ 
    & 25  & 0.47 (0.09) & 0.43 (0.09) & 0.5 (0.09) & 0.53 (0.06) & 0.57 (0.05) & 0.51 (0.06) & 127.17 (22.22) & 131.43 (18.97) & 134.47 (17.22) & 132.42 (18.99) \\ 
    & 50  & 0.5 (0.09) & 0.43 (0.09) & 0.5 (0.09) & 0.48 (0.06) & 0.56 (0.04) & 0.48 (0.05) & 125.72 (21.8) & 129.27 (18.16) & 129.59 (17.59) & 130.27 (18.15) \\ 
    & 100  & 0.47 (0.09) & 0.47 (0.09) & 0.43 (0.09) & 0.43 (0.06) & 0.57 (0.04) & 0.47 (0.06) & 128.38 (21.86) & 130.05 (18.05) & 131.83 (19.08) & 130.33 (18.06) \\
\end{tabular}}
   \end{center}
\end{table}
\end{landscape}

\begin{landscape}
\begin{table}[b]
\caption{30 Monte Carlo simulations of $\mc{M}_2$ for $\mc{B}_f$ with varying $n$ and $\sigma_{\alpha}$} \vspace{.3cm} \label{table3}
\begin{center}
\resizebox{1.4\textwidth}{!}{\begin{tabular}{cc|ccc|ccc|cccc|}
   &   & \multicolumn{3}{|c|}{Guess} & \multicolumn{3}{|c|}{LOOCV with $k$ random draws} &  \multicolumn{4}{|c|}{Distance to $y_{1, T_1^*+1}$} \\ 
 $n$   & $\sigma_{\alpha}$ &  $\delta_{\hat{\alpha}_{\rm adj}} $  & $\delta_{\hat{\alpha}_{\rm wadj}}$ & $\delta_{\hat{\alpha}_{\rm IVW}}$  & $\bar{\mc{C}}^{(k)}(\delta_{\hat{\alpha}_{\rm adj}})$  & $\bar{\mc{C}}^{(k)}(\delta_{\hat{\alpha}_{\rm wadj}})$ & $\bar{\mc{C}}^{(k)}(\delta_{\hat{\alpha}_{\rm IVW}})$ & Original & $\hat{\alpha}_{\rm adj}$ & $\hat{\alpha}_{\rm wadj}$ & $\hat{\alpha}_{\rm IVW}$\\[.15cm] 
  \hline
 \multirow{5}{*}{5} & 5  & 1 (0) & 1 (0) & 1 (0) & 0.89 (0.03) & 0.92 (0.02) & 0.89 (0.03) & 48.52 (3.93) & 15.74 (2.34) & 15.76 (2.34) & 15.16 (2.24) \\ 
  & 10  & 1 (0) & 1 (0) & 1 (0) & 0.89 (0.02) & 0.91 (0.02) & 0.89 (0.02) & 47.7 (4.35) & 18.26 (2.37) & 18.97 (2.42) & 17.68 (2.28) \\ 
  & 25  & 0.97 (0.03) & 1 (0) & 0.93 (0.05) & 0.79 (0.03) & 0.81 (0.03) & 0.77 (0.03) & 46.95 (6.11) & 27.35 (3.81) & 30.88 (3.84) & 26.58 (3.83) \\ 
  & 50  & 0.8 (0.07) & 0.93 (0.05) & 0.8 (0.07) & 0.62 (0.03) & 0.65 (0.03) & 0.63 (0.03) & 56.85 (8.64) & 46.96 (7.02) & 52.92 (7.45) & 45.95 (7.17) \\ 
  & 100  & 0.73 (0.08) & 1 (0) & 0.8 (0.07) & 0.53 (0.04) & 0.53 (0.04) & 0.55 (0.04) & 99.22 (12.84) & 93.4 (12.97) & 103.82 (14.1) & 91.95 (13.34) \\[.3cm]  
 \multirow{5}{*}{10} & 5  & 1 (0) & 1 (0) & 1 (0) & 0.86 (0.03) & 0.88 (0.02) & 0.86 (0.03) & 50.59 (5.24) & 29.19 (5.2) & 31.4 (5.28) & 29.29 (5.22) \\ 
  & 10  & 1 (0) & 1 (0) & 1 (0) & 0.82 (0.03) & 0.84 (0.03) & 0.82 (0.03) & 51.17 (5.49) & 31.55 (5.33) & 33.91 (5.56) & 31.7 (5.35) \\ 
  & 25  & 0.93 (0.05) & 1 (0) & 0.93 (0.05) & 0.72 (0.04) & 0.75 (0.04) & 0.71 (0.04) & 53.53 (6.58) & 40.05 (6.03) & 43.5 (6.66) & 40.43 (6) \\ 
  & 50  & 0.87 (0.06) & 0.97 (0.03) & 0.87 (0.06) & 0.55 (0.04) & 0.58 (0.05) & 0.55 (0.05) & 62.45 (8.25) & 55.56 (8.11) & 62.15 (9.19) & 56.12 (8.04) \\ 
  & 100  & 0.77 (0.08) & 0.97 (0.03) & 0.73 (0.08) & 0.49 (0.05) & 0.44 (0.05) & 0.46 (0.05) & 85.72 (12.73) & 89.5 (13.37) & 103.25 (15.07) & 89.92 (13.29) \\[.3cm] 
 \multirow{5}{*}{15} & 5  & 1 (0) & 1 (0) & 1 (0) & 0.95 (0.02) & 0.92 (0.03) & 0.95 (0.02) & 52.1 (2.96) & 14.04 (1.78) & 13.36 (2.07) & 14.11 (1.76) \\ 
  & 10  & 1 (0) & 1 (0) & 1 (0) & 0.92 (0.02) & 0.9 (0.03) & 0.93 (0.02) & 52.25 (3.3) & 15.12 (1.93) & 14.24 (2.29) & 15.18 (1.9) \\ 
    & 25  & 0.93 (0.05) & 1 (0) & 0.9 (0.06) & 0.8 (0.03) & 0.8 (0.03) & 0.8 (0.03) & 52.71 (5.28) & 22.98 (2.9) & 22.6 (3.42) & 22.95 (2.88) \\ 
    & 50  & 0.7 (0.09) & 0.9 (0.06) & 0.7 (0.09) & 0.65 (0.03) & 0.65 (0.03) & 0.65 (0.04) & 58.65 (8.48) & 39.51 (5.65) & 40.8 (6.3) & 39.35 (5.63) \\ 
    & 100  & 0.6 (0.09) & 0.87 (0.06) & 0.6 (0.09) & 0.47 (0.05) & 0.45 (0.04) & 0.45 (0.05) & 88.76 (13.66) & 75.93 (11.52) & 81.89 (12.19) & 75.94 (11.41) \\[.3cm] 
   \multirow{5}{*}{25} & 5  & 1 (0) & 1 (0) & 1 (0) & 0.94 (0.02) & 0.95 (0.02) & 0.94 (0.02) & 50.55 (2.9) & 12.13 (1.77) & 14.22 (1.96) & 12.09 (1.77) \\ 
    & 10  & 1 (0) & 1 (0) & 1 (0) & 0.93 (0.02) & 0.95 (0.02) & 0.93 (0.02) & 49.13 (3.31) & 14.78 (1.85) & 18.21 (2) & 14.75 (1.85) \\ 
    & 25  & 1 (0) & 1 (0) & 1 (0) & 0.83 (0.02) & 0.85 (0.02) & 0.83 (0.03) & 47.38 (5.13) & 26.85 (3.33) & 32.95 (3.59) & 26.81 (3.32) \\ 
    & 50  & 0.97 (0.03) & 1 (0) & 0.97 (0.03) & 0.61 (0.04) & 0.71 (0.04) & 0.62 (0.04) & 56.73 (7.93) & 50.96 (6.63) & 60.62 (7.21) & 50.88 (6.59) \\ 
    & 100  & 0.8 (0.07) & 0.93 (0.05) & 0.8 (0.07) & 0.49 (0.05) & 0.51 (0.04) & 0.49 (0.05) & 93.79 (14.67) & 102.05 (13.37) & 116.45 (15.19) & 101.55 (13.38) \\
\end{tabular}}
   \end{center}
      \vspace{-.5cm}
\end{table}
\end{landscape}

\begin{landscape}
\begin{table}[b]
\caption{30 Monte Carlo simulations of $\mc{M}_2$ for $\mc{B}_f$ with varying $\sigma$ and $\sigma_{\alpha}$} \vspace{.3cm} \label{table4}
\begin{center}
\resizebox{1.4\textwidth}{!}{\begin{tabular}{cc|ccc|ccc|cccc|}
   &   & \multicolumn{3}{|c|}{Guess} & \multicolumn{3}{|c|}{LOOCV with $k$ random draws} &  \multicolumn{4}{|c|}{Distance to $y_{1, T_1^*+1}$} \\ 
 $\sigma$   & $\sigma_{\alpha}$ &  $\delta_{\hat{\alpha}_{\rm adj}}$  & $\delta_{\hat{\alpha}_{\rm wadj}}$ & $\delta_{\hat{\alpha}_{\rm IVW}}$  & $\bar{\mc{C}}^{(k)}(\delta_{\hat{\alpha}_{\rm adj}})$  & $\bar{\mc{C}}^{(k)}(\delta_{\hat{\alpha}_{\rm wadj}})$ & $\bar{\mc{C}}^{(k)}(\delta_{\hat{\alpha}_{\rm IVW}})$ & Original & $\hat{\alpha}_{\rm adj}$ & $\hat{\alpha}_{\rm wadj}$ & $\hat{\alpha}_{\rm IVW}$\\[.15cm] 
  \hline
 \multirow{5}{*}{5} & 5  & 1 (0) & 1 (0) & 1 (0) & 0.94 (0.02) & 0.95 (0.02) & 0.94 (0.02) & 50.04 (3.65) & 21.75 (4.34) & 22.68 (4.41) & 21.83 (4.34) \\ 
  & 10  & 1 (0) & 1 (0) & 1 (0) & 0.93 (0.02) & 0.92 (0.02) & 0.93 (0.02) & 49.93 (4.11) & 24.3 (4.44) & 25.72 (4.58) & 24.4 (4.45) \\ 
  & 25  & 0.97 (0.03) & 1 (0) & 0.97 (0.03) & 0.73 (0.04) & 0.8 (0.03) & 0.73 (0.04) & 51.29 (5.39) & 32.8 (5.25) & 35.39 (5.78) & 33.11 (5.21) \\ 
  & 100  & 0.77 (0.08) & 0.97 (0.03) & 0.7 (0.09) & 0.49 (0.04) & 0.44 (0.05) & 0.47 (0.04) & 82.55 (11.61) & 84.26 (12.7) & 98.66 (13.96) & 84.62 (12.59) \\[.3cm] 
 \multirow{5}{*}{10} & 5  & 1 (0) & 1 (0) & 1 (0) & 0.86 (0.03) & 0.88 (0.02) & 0.86 (0.03) & 50.59 (5.24) & 29.19 (5.2) & 31.4 (5.28) & 29.29 (5.22) \\ 
  & 10  & 1 (0) & 1 (0) & 1 (0) & 0.82 (0.03) & 0.84 (0.03) & 0.82 (0.03) & 51.17 (5.49) & 31.55 (5.33) & 33.91 (5.56) & 31.7 (5.35) \\ 
  & 25  & 0.93 (0.05) & 1 (0) & 0.93 (0.05) & 0.72 (0.04) & 0.75 (0.04) & 0.71 (0.04) & 53.53 (6.58) & 40.05 (6.03) & 43.5 (6.66) & 40.43 (6) \\ 
  & 50  & 0.87 (0.06) & 0.97 (0.03) & 0.87 (0.06) & 0.55 (0.04) & 0.58 (0.05) & 0.55 (0.05) & 62.45 (8.25) & 55.56 (8.11) & 62.15 (9.19) & 56.12 (8.04) \\ 
  & 100  & 0.77 (0.08) & 0.97 (0.03) & 0.73 (0.08) & 0.49 (0.05) & 0.44 (0.05) & 0.46 (0.05) & 85.72 (12.73) & 89.5 (13.37) & 103.25 (15.07) & 89.92 (13.29)\\[.3cm] 
 \multirow{5}{*}{25} & 5  & 0.97 (0.03) & 1 (0) & 0.97 (0.03) & 0.7 (0.03) & 0.73 (0.03) & 0.71 (0.04) & 57.87 (8.76) & 50.31 (7.58) & 57.25 (7.75) & 50.53 (7.64) \\ 
  & 10  & 0.97 (0.03) & 1 (0) & 0.97 (0.03) & 0.68 (0.04) & 0.69 (0.04) & 0.69 (0.04) & 58.41 (9.11) & 51.62 (7.91) & 58.61 (8.28) & 51.85 (7.98) \\ 
  & 25  & 0.93 (0.05) & 0.97 (0.03) & 0.9 (0.06) & 0.63 (0.04) & 0.68 (0.04) & 0.63 (0.04) & 62.02 (10.02) & 59.08 (8.53) & 65.55 (9.64) & 59.68 (8.53) \\ 
  & 50  & 0.87 (0.06) & 0.9 (0.06) & 0.87 (0.06) & 0.54 (0.04) & 0.59 (0.04) & 0.52 (0.04) & 71.01 (11.73) & 73.52 (10.18) & 81.81 (11.98) & 74.51 (10.09) \\ 
  & 100  & 0.73 (0.08) & 0.87 (0.06) & 0.73 (0.08) & 0.47 (0.05) & 0.45 (0.04) & 0.5 (0.05) & 95.93 (15.59) & 104.49 (15.12) & 119.38 (17.62) & 105.72 (14.97) \\[.3cm] 
 \multirow{5}{*}{50} & 5  & 0.8 (0.07) & 0.77 (0.08) & 0.8 (0.07) & 0.52 (0.04) & 0.49 (0.05) & 0.52 (0.04) & 85.95 (14.57) & 90.02 (13.45) & 103.03 (13.91) & 90.08 (13.64) \\ 
  & 10  & 0.8 (0.07) & 0.73 (0.08) & 0.8 (0.07) & 0.55 (0.05) & 0.5 (0.04) & 0.53 (0.05) & 86.44 (14.95) & 90.89 (13.79) & 104.62 (14.28) & 91.03 (13.98) \\ 
  & 25  & 0.77 (0.08) & 0.77 (0.08) & 0.77 (0.08) & 0.53 (0.04) & 0.46 (0.04) & 0.53 (0.04) & 90.26 (15.82) & 95.36 (14.69) & 109.53 (15.69) & 95.79 (14.83) \\ 
  & 50  & 0.77 (0.08) & 0.8 (0.07) & 0.77 (0.08) & 0.48 (0.05) & 0.45 (0.05) & 0.45 (0.05) & 99.52 (17.26) & 106.52 (16.17) & 120.99 (18.13) & 107.55 (16.19) \\ 
  & 100  & 0.57 (0.09) & 0.77 (0.08) & 0.63 (0.09) & 0.41 (0.04) & 0.45 (0.03) & 0.41 (0.04) & 123.11 (20.61) & 135.24 (19.74) & 151.62 (23.42) & 137.21 (19.55) \\[.3cm] 
 \multirow{5}{*}{100} & 5  & 0.63 (0.09) & 0.57 (0.09) & 0.63 (0.09) & 0.48 (0.05) & 0.48 (0.03) & 0.47 (0.05) & 156.82 (26.36) & 170.06 (25.92) & 196.4 (26.55) & 170.13 (26.27) \\ 
  & 10  & 0.63 (0.09) & 0.57 (0.09) & 0.67 (0.09) & 0.46 (0.05) & 0.47 (0.03) & 0.47 (0.05) & 157.3 (26.76) & 170.93 (26.21) & 197.96 (26.87) & 171.07 (26.56) \\ 
  & 25  & 0.67 (0.09) & 0.63 (0.09) & 0.67 (0.09) & 0.44 (0.04) & 0.5 (0.03) & 0.45 (0.04) & 160.32 (27.73) & 173.66 (27.22) & 202.61 (28.01) & 174.28 (27.48) \\ 
  & 50  & 0.67 (0.09) & 0.67 (0.09) & 0.67 (0.09) & 0.39 (0.04) & 0.43 (0.03) & 0.38 (0.04) & 166.98 (29.35) & 182.83 (28.37) & 210.38 (30.48) & 183.67 (28.61) \\ 
  & 100  & 0.6 (0.09) & 0.67 (0.09) & 0.53 (0.09) & 0.4 (0.04) & 0.45 (0.04) & 0.41 (0.05) & 188.29 (32.07) & 203.7 (31.63) & 233.44 (35.31) & 205.4 (31.72) \\
\end{tabular}}
   \end{center}
\end{table}
\end{landscape}

\bibliographystyle{plainnat}
\bibliography{synthetic-prediction-notes}

\end{document}